\documentclass[a4paper,12pt]{article}
\usepackage[utf8]{inputenc}
\usepackage{jheppub}
\usepackage{tikz}
\usepackage[miktex]{gnuplottex}
\usetikzlibrary{shapes,arrows,positioning,automata,backgrounds,calc,er,patterns}
\usepackage{tikz-feynman}
\tikzfeynmanset{compat=1.0.0}
\usepackage{soul}
\usepackage{slashed} 
\usepackage{enumerate}
\usepackage{epsfig} 
\usepackage{float} 
\usepackage{subcaption} 
\usepackage{amsmath} 
\usepackage{wasysym} 
\usepackage{mathrsfs} 
\usepackage{amsfonts} 
\usepackage{amsbsy} 
\usepackage{amscd}    
\usepackage{multirow} 
\usepackage{tikz}
\usepackage{slashed}
\usepackage[miktex]{gnuplottex}
\usetikzlibrary{arrows,positioning,shapes.geometric} 
\usepackage{color}
\usepackage{verbatim}
\usepackage{xcolor}
\usepackage{aligned-overset}

\makeatletter

\def\eeq{\end{equation}}
\def\beq{\begin{equation}}

\newcommand{\Rmnum}[1]{\expandafter\@slowromancap\romannumeral #1@}

\newcommand{\cbra}[1]{\left\{#1\right\}}
\newcommand{\rbra}[1]{\left(#1\right)}
	\newcommand{\sbra}[1]{\left[#1\right]}

\newcommand{\bea} {\begin{eqnarray}}
\newcommand{\eea} {\end{eqnarray}}

	\newcommand{\nn}{\nonumber}

	\newcommand{\re}{\textrm{Re}}
	\newcommand{\im}{\textrm{Im}}

\makeatother

\title{CP-violation and its implications in a complex singlet extension of 2HDM}
\author[c]{Jayita Lahiri,}
  
\author[a,b]{Gudrid Moortgat-Pick}

\affiliation[a]{II. Institut f{\"u}r Theoretische Physik, Universit{\"a}t Hamburg, Luruper Chaussee 149, 22761 Hamburg, Germany} 

\affiliation[b]{Deutsches Elektronen-Synchrotron DESY, Notkestr.\ 85, 22607 Hamburg, Germany}

\affiliation[c]{Department of Physical Sciences, Indian Institute of Science Education and Research Kolkata, Mohanpur - 741246, India}

\emailAdd{jayita.lahiri@desy.de}
\emailAdd{gudrid.moortgat-pick@desy.de}

\abstract{ We investigate CP-violation in the complex singlet extension of the general Two Higgs Doublet Model (2HDM) with Yukawa alignment condition. We first explore the possibility of explicit CP-violation in the extended scalar sector while the 125 GeV Higgs remains exactly Standard Model (SM)-like. We identify an additional source of CP violation in the complex singlet extension compared to the 2HDM, which allows this model a substantially greater freedom in satisfying the stringent EDM constraints. We also incorporate dark matter in this model and investigate the impacts of constraints from the dark sector on the model parameter space and its interplay with CP-violation phases. We further explore the possibility of detecting such a scenario at future collider experiments via CP-violating trilinear couplings among the non-standard scalars. Finally, we also deviate from the exact alignment limit and investigate the CP-properties of the observed Higgs boson in the context of our model. We demonstrate the strong model-dependent nature of the detection prospects of the CP-phase of the Higgs boson at future experiments, exploring both fermion couplings as well as the trilinear self-coupling of the Higgs boson. 

}

\preprint{DESY-26-035}
\begin{document}

\maketitle

\newpage

\section{Introduction}
\label{sec1}

The Standard Model (SM) of particle physics has been undergoing continuous experimental scrutiny, confirming that it is an extremely successful `effective' theory.
However, there are still unanswered questions that need attention. One of such questions is 
the explanation of the matter-antimatter asymmetry of the universe.
The dominance of matter over antimatter has puzzled us for a long time. Quantitatively, this asymmetry can be written as 

\begin{equation}
 \eta=  \frac{n_B - n_{\bar B}}{n_{\gamma}} = 6 \times 10^{-10}, 
\end{equation}

\noindent
where $n_B$ is the number of baryons, $n_{\bar B}$ the number of anti-baryons and $n_\gamma$ is the number of photons in the universe.
The process, where the observed non-vanishing baryon asymmetry of the universe (BAU) was produced, is known as baryogenesis. For a successful baryogenesis, the three Sakharov conditions need to be satisfied~\cite{Sakharov:1967dj}:
\begin{itemize}
\item Baryon number violation,
\item C and CP-violation,
\item Departure from thermal equilibrium. 
\end{itemize}

\noindent
In the Standard Model, the elements mentioned above are present: for example, baryon number conservation is violated in the sphaleron process, the weak interaction violates C (charge conjugation), and CP-violation is also present via the phases of Cabbibo-Kobayashi-Maskawa (CKM) matrix. But the predicted CP-violating effects are too small, and therefore, the numerical value of the observed BAU is also too low within the SM. Furthermore, a strong first-order phase transition, which is an out-of-equilibrium process, is not achievable within the SM. 
Therefore, in order to have a successful baryogenesis, we need to go beyond the SM. In the past, various theories have been proposed, such as baryogenesis with Grand Unified Theories (GUT baryogenesis) \cite{Yoshimura:1978ex,Weinberg:1979bt}, baryogenesis via leptogenesis ~\cite{Fukugita:1986hr,Luty:1992un}, and electroweak baryogenesis ~\cite{Kuzmin:1985mm,Shaposhnikov:1987tw,Trodden:1998ym,Morrissey:2012db}. A large number of studies have been done in the context of electroweak baryogenesis, in models with an extended scalar sector, for instance in the two Higgs Doublet Model (2HDM) and its extensions~\cite{Fromme:2006cm,Profumo:2007wc,Arhrib:2010ju,Cline:2011mm,Biekotter:2025fjx,Biekotter:2025vxl}. 

While CP-violation beyond the SM is a major motivation for our work, another strong motivation comes from the existence of dark matter (DM). There is astrophysical and cosmological evidence of dark matter as, for instance, in the rotation curves of galaxies or the gravitational lensing effects. But the SM does not provide a candidate for a so-called cold DM candidate. In order to accommodate a DM candidate in the theory, one needs to go beyond the SM, and there are interesting possibilities of introducing a scalar dark matter in various scalar extensions of the SM. Dark matter phenomenology has been studied, for instance, in the scalar extensions of the 2HDM~\cite{Dey:2019lyr,Dey:2021alu,Dutta:2023cig,Dutta:2025nmy},  considering 
the interesting possibility of non-standard scalars acting as a portal to the dark sector.

It is interesting to ask at this point whether one can obtain sufficient CP-violation and at the same time, accommodate a stable dark matter candidate within such models.
Alongside addressing both the CP-violation and the dark matter problem, a complex singlet extension of the 2HDM has one interesting feature: it was pointed out in~\cite{Heinemeyer:2021msz,Dutta:2023cig}, that it is possible to explain the recently observed 95 GeV excess in such models. In this work, however, we do not focus on accommodating a 95 GeV excess within this model, but we study non-standard scalars of moderate mass range (200-400)GeV, that are allowed by all constraints.

While it is interesting to see that several questions can be answered at the same time within this model, one should also remember that there are numerous constraints on such scenarios. For example, CP-violation will also lead to the electric dipole moments (EDM) of the electron and neutron, which are precisely measured in the experiments, especially in the case of the electron EDM.  Since we focus on CP-violation, we must take into account constraints from these experiments. 

There have been studies in the past, where CP-violation has been explored within an Effective Field Theory (EFT) approach, {\it i.e.} without going into the details of any specific model~\cite{Li:2025ouv,ATLAS:2020ior}. We take a different approach for the current study, motivated by the following reasons. 

\begin{itemize}
\item Our final goal is to probe CP-violation at future collider experiments, for which we must know the size of CP-violating phases precisely, which are allowed by the EDM experiments and which can be probed at future collider experiments. 
Such constraints from EDM on CP-violating phases are, in general, model-dependent. Therefore, the prospect of probing them at future colliders would also be model-dependent and would not be appropriate to be predicted only by using an effective approach. For this purpose, we need a model-dependent, detailed analysis. 
\item The SMEFT approach usually focuses on the CP-violation in the 125 GeV SM-like Higgs boson. However, such a CP-phase is quite strongly constrained by the data, and it is possible that significant CP-violation is lurking in the BSM Higgs sector, such a scenario is impossible to probe in the SMEFT approach, and requires a specific model description of the extended scalar sector. 
\end{itemize}

\noindent
Furthermore, we investigate, in particular, the impact of CP-violating parameters on the dark matter phenomenology. We also consider theoretical constraints and experimental constraints such as bounds from the Higgs data, from direct searches of heavy scalars at the LHC and from B-physics on the model parameter space. 

\noindent
In section~\ref{model}, we discuss the details of our model, sources of CP-violating phases and the dark matter sector. In section~\ref{constraints}, we discuss the constraints coming from EDM experiments, dark matter observables and existing theoretical and experimental constraints mentioned above. We discuss the prospect of discovering CP-violation in the BSM sector via certain final states in section~\ref{collider}. In section~\ref{collider125}, we investigate the CP-phase of the 125 GeV Higgs boson, allowed by the EDM bounds, and that can be probed at the future colliders. Finally, we summarize our results and conclude in section~\ref{conclusion}.

\section{The model - complex singlet extension of 2HDM with dark matter }
\label{model}
We consider the most general potential involving two Higgs doublets + a complex singlet. The scalar sector of such a model consists of two scalar doublets $\Phi_1$ and $\Phi_2$, both with hypercharge $Y=1$, same as in 2HDM. The complex singlet extension contains, in addition, a complex singlet field $\Phi_S$ with

\begin{equation*}
V_{\text{2HDMS}} = V_{\text{2HDM}} + V_S,
\end{equation*}

\noindent
where $V_{\text{2HDM}}$ is the most general 2HDM potential given below. $V_S$, on the other hand, encapsulates the self-interaction of the complex singlet field as well as the interaction between the two doublets and the complex singlet:

\begin{eqnarray}
 \label{eq:2HDM_potential}V_{2HDM} &=& -m_{11}^2 \Phi_1^{\dagger} \Phi_1 - m_{22}^2 \Phi_2^{\dagger} \Phi_2 - [m_{12}^2 \Phi_1^{\dagger} \Phi_2 + h.c. ] + \frac{\lambda_1}{2} (\Phi_1^{\dagger} \Phi_1)^2 \nonumber\\ 
     &+& \frac{\lambda_2}{2}(\Phi_2^{\dagger} \Phi_2)^2  + \lambda_3 (\Phi_1^{\dagger} \Phi_1) (\Phi_2^{\dagger} \Phi_2) + \lambda_4 (\Phi_1^{\dagger} \Phi_2) (\Phi_2^{\dagger} \Phi_1) \nonumber\\
   &+&  \left[\frac{\lambda_5}{2} (\Phi_1^{\dagger} \Phi_2) + \lambda_6 (\Phi_1^{\dagger} \Phi_1) + \lambda_7 (\Phi_2^{\dagger} \Phi_2)\right] (\Phi_1^{\dagger} \Phi_2) +h.c.,  
\end{eqnarray}

\begin{eqnarray}
V_{S} &=& m_S^2 \Phi_S^{\dagger} \Phi_S + \left[ \frac{m_S'^2}{2} \Phi_S^2 + h.c. \right] + \left[ \frac{\lambda_1''}{24} \Phi_S^4 + h.c. \right] + \left[ \frac{\lambda_2''}{6} (\Phi_S^2 \Phi_S^{\dagger} \Phi_S) + h.c. \right]\nonumber \\
  &+& \frac{\lambda_3''}{4}(\Phi_S^{\dagger} \Phi_S)^2 + \Phi_S^{\dagger} \Phi_S [\lambda_1' \Phi_1^{\dagger} \Phi_1 + \lambda_2' \Phi_2^{\dagger} \Phi_2] + [\Phi_S^2(\lambda_4' \Phi_1^{\dagger} \Phi_1 + \lambda_5' \Phi_2^{\dagger} \Phi_2) + h.c.] \nonumber \\
 &+& [\lambda_6' \Phi_1^{\dagger} \Phi_2 \Phi_S^{\dagger} \Phi_S + h.c] + [\lambda_7' \Phi_1^{\dagger} \Phi_2 \Phi_S^2 + h.c] + [\lambda_8' \Phi_2^{\dagger} \Phi_1 \Phi_S^2 + h.c].
 \label{2hdms}
\end{eqnarray}


\noindent
One can see that the coefficients $m_{12}^2, \lambda_5, \lambda_6, \lambda_7$ in the general 2HDM potential can be complex. Furthermore $m_S'^2, \lambda_1'', \lambda_2'', \lambda_4', \lambda_5', \lambda_6', \lambda_7', \lambda_8' $ in the complex singlet interaction potential $V_S$ can also be complex. The complex coefficients hint at the possibility of CP-violation in the scalar sector. However, it is well-known that the 2HDM potential respects certain symmetries. For example, the 2HDM transformation is symmetric under a global U(2) transformation. Under this symmetry, some of the complex phases can be rotated away, and therefore, not all the complex phases are physical. A symmetry transformation can make the unphysical phases go away.  

In the context of 2HDM, it is customary to introduce a $Z_2$-symmetry on the Lagrangian, in order to suppress the tree-level flavor-changing neutral current (FCNC). If one assumes the $Z_2$-symmetry is exact, {\it i.e.} $m_{12}^2=\lambda_6=\lambda_7=0$, (Inert Doublet model), in that case the only remaining complex phase of $\lambda_5$ can be rotated away by the global $U(2)$ transformation~\cite{Ginzburg:2004vp}. Therefore, there cannot be any CP-violation. If the $Z_2$ symmetry is softly broken, which means $m_{12}^2\neq 0$, $\lambda_6=\lambda_7=0$, this condition is sufficient to ensure suppression of tree-level FCNC. It has been shown in~\cite{Gunion:2005ja,Grzadkowski:2014ada,Grzadkowski:2016szj} that, in this case, there can be CP-violation in the model. But the CP-violation becomes negligible when we impose the `alignment' condition {\it i.e.} one of the scalars in the model corresponds to the 125 GeV Standard Model-like Higgs boson. In other words, the imaginary part of the U(2) invariants constructed out of the scalar potential vanishes in the exact alignment limit. In \cite{Grzadkowski:2014ada}, the basis-independent quantities were expressed in terms of the physical observables such as masses and couplings. It has been shown in~\cite{Grzadkowski:2016szj} that in the exact alignment limit, both explicit and spontaneous CP-violation vanish in the softly broken $Z_2$-symmetric 2HDM. 

Therefore, it is imperative to go beyond the $Z_2$-symmetric 2HDM, {\it i.e.} $\lambda_6,\lambda_7 \neq 0$ in order to introduce CP-violation in the exact alignment limit. But in the absence of the softly broken $Z_2$-symmetry, we run into the danger of introducing tree-level FCNC in the model. In an earlier work~\cite{Pich:2009sp}, it was shown that in the absence of softly broken $Z_2$-symmetry, it is possible to circumvent the tree-level FCNC, if the two Yukawa matrices corresponding to fermion couplings to the two doublets are proportional to each other. In that case, the two matrices can be simultaneously diagonalized and tree-level FCNC can be avoided.
Such a scenario is called Yukawa-aligned 2HDM. The Yukawa sector remains the same if we extend the Yukawa-aligned 2HDM with a complex-singlet extension. We will shortly discuss the Yukawa sector in detail.

\subsection{Conditions for dark matter in the model}

Next, we investigate whether it is possible to accommodate a dark matter candidate in this model. In a complex singlet extension of 2HDM with real parameters~\cite{Dutta:2023cig,Dutta:2025nmy}, the potential in Eq.~\ref{eq:2HDM_potential} and \ref{2hdms} would possess a ${\cal{Z}}_2 \times {\cal{Z}}_2'$ symmetry. The real parameters ensure cancellation between terms breaking this symmetry (terms involving cubic terms) and their Hermitian conjugates.
If one of the discrete symmetries of the ${\cal{Z}}_2 \times {\cal{Z}}_2'$ symmetry is broken spontaneously, while the other remains intact, {\it i.e.} $\Phi_S = v_S + h_S + i a_S$, then $a_S$ can act as a dark matter candidate. We have explored this scenario in our earlier works~\cite{Dutta:2023cig,Dutta:2025nmy}.

The situation is quite different when the potential in Eqs.~\ref{eq:2HDM_potential} and \ref{2hdms} contains complex parameters. In the absence of the aforementioned cancellation, as in the real case, the ${\cal{Z}}_2 \times {\cal{Z}}_2'$ breaking terms remain in the potential. Under such circumstances, achieving a dark matter candidate that will neither mix with other scalar degrees of freedom, nor decay into scalar states, is not straightforward, but still possible to achieve. This would require imposing certain conditions on the parameters of the potential.

The list of bilinear terms that can introduce forbidden mixing between DM state $a_S$ and other scalar states,  as well as the trilinear and quartic couplings, leading to two-body and three-body decays of DM, is given in Appendix~\ref{dm_interactions}. It follows from those relations that
the necessary conditions for a stable dark matter in this model are as follows:

 The parameters $\lambda_4'$, $\lambda_5'$, $m_S'^2$ are real and
 \begin{eqnarray}
   \text{Re}[\lambda_7'] &=& \text{Re}[\lambda_8'], \nonumber \\
      \text{Im}[\lambda_7'] &=& -\text{Im}[\lambda_8'], \nonumber \\
   \text{Im}[\lambda_1''] &=& -2\times \text{Im}[\lambda_2''] 
   \label{dm}.
\end{eqnarray}

\noindent
In this case, we will be left with three independent phases, of $\lambda_6'$, $\lambda_7'$ and $\lambda_1''$.

\subsection{Mass matrix and CP-violating phase in the scalar sector }

For the purpose of simplicity, we work in the Higgs basis defined as follows. The relation between the parameters in the Higgs basis and those in the interaction basis in this model is given in Appendix~\ref{basis_change}.

\begin{equation*}
\Phi_1=
\left(
\begin{array}{c} 
  G^+ \\ 
  \frac{1}{\sqrt{2}}(v+h_1 + iG^0)
\end{array} 
\right),~~\Phi_2=
\left(
\begin{array}{c} 
  H^+ \\ 
  \frac{1}{\sqrt{2}}(h_2 + i a_2)
\end{array} 
\right),~~\Phi_S = v_S + h_S + i a_S.
\end{equation*}

\noindent
In the Higgs basis, the minimization conditions for the scalar potential are as follows. The following mass parameters can then be substituted with vev's in the scalar potential.

\begin{eqnarray}
m_{11}^2 &=& \frac{1}{2} \lambda_1 v^2 + \frac{1}{2} \lambda_1' v_S^2 + \text{Re}[\lambda_4'] v_S^2, \nonumber\\
\text{Re}[m_{12}^2] &=& \frac{1}{2} (\text{Re}[\lambda_6] v^2 + \text{Re}[\lambda_6'] v_S^2 + \text{Re}[\lambda_7'] v_S^2 + \text{Re}[\lambda_8'] v_S^2),\nonumber \\
\text{Im}[m_{12}^2] &=& \frac{1}{2} (\text{Im}[\lambda_6] v^2 + \text{Im}[\lambda_6'] v_S^2 + \text{Im}[\lambda_7'] v_S^2 - \text{Im}[\lambda_8'] v_S^2),\nonumber \\
m_S^2 &=& -(\text{Re}[m_S'^2] + \frac{1}{2} \lambda_1' v^2 + \text{Re}[\lambda_4'] v^2) + \left(\frac{\text{Re}[\lambda_1'']}{12} + \frac{\text{Re}[\lambda_2'']}{3} + \frac{\text{Re}[\lambda_3'']}{4}\right)v_S^2,\nonumber \\
\text{Im}[m_S'^2] &=& -\left(\frac{\text{Im}[\lambda_1'']}{12} + \frac{\text{Im}[\lambda_2'']}{6}\right)v_S^2 + \text{Im}[\lambda_4']v^2.
\label{minimization}
\end{eqnarray}

\noindent
Furthermore, dark Matter mass can be obtained as follows:
\begin{equation}
m_{\text{DM}}^2 = -2 \text{Re}[m_S'^2] - \frac{1}{3} v_S^2 (\text{Re}[\lambda_1''] + \text{Re}[\lambda_2'']) - 
   2 v^2 \text{Re}[\lambda_4'].
\end{equation}

\noindent
With the assumption of an exact alignment condition and after demanding a stable dark matter candidate,
the mass matrix for the neutral scalars in the Higgs basis ($h_1, h_2, a_2, h_S, a_S$) is as follows:

\begin{equation}
{\cal{M}}_{ij}^2=
\left(
\begin{array}{c|ccc|c} 
  M_{11} & 0 & 0 & 0 & 0 \nonumber \\ 
  \hline 
  0 & M_{22} & 0 & M_{24} & 0 \nonumber\\
  0 & 0 & M_{33} & M_{34} & 0 \nonumber\\
  0 & M_{24} & M_{34} & M_{44} & 0 \nonumber\\
  \hline  
  0 & 0 & 0 & 0 & m_{55} \\
\end{array} 
\right).
\end{equation}

\noindent
The zero elements in the first row and first column of the mass matrix in ${\cal{M}}_{ij}^2$ are indicative of the fact that the Higgs boson is exactly SM-like, {\it i.e.} the alignment condition is exact. The alignment condition is ensured by the following conditions.

\begin{eqnarray}
\text{Re}[\lambda_6] &=& 0, \nonumber \\
\text{Im}[\lambda_6] &=& 0, \nonumber \\
\text{Re}[\lambda_1'] &=& -2\times\text{Re}[\lambda_4'].
\label{alignment}
\end{eqnarray}

\noindent
The zero element in the fifth row and column is also a result of having a dark matter candidate, which does not mix with any other scalars in the theory. We reiterate that this feature is ensured by Eqs.~\ref{dm}.

Keeping these in mind, one can rotate the eigenstates in the Higgs basis $(h_1, h_2, a_2, h_S, a_S)$ to the mass basis to get the mass eigenstates $(H_1, H_2, H_3, H_4, \text{DM})$. The block-diagonal structure of the mass matrix $M_{ij}^2$ tells us that the rotation matrix $R_{ij}$ will also be block-diagonal: 

 \begin{equation}
{\cal{R}}_{ij}=
\left(
\begin{array}{c|ccc|c} 
  1 & 0 & 0 & 0 & 0 \nonumber \\ 
  \hline 
  0 & R_{22} & R_{23} & R_{24} & 0 \nonumber\\
  0 & R_{32} & R_{33} & R_{34} & 0 \nonumber\\
  0 & R_{42} & R_{43} & R_{44} & 0 \nonumber\\
  \hline  
  0 & 0 & 0 & 0 & 1 \\
\end{array} 
\right).
\end{equation}

\noindent
We can see that the block diagonal nature of $R_{ij}$ ensures no mixing between the SM-like Higgs ($h_1$) and non-standard scalars {\it i.e.} $H_1=h_1$ and decoupling of the DM from the visible sector, {\it i.e.} $a_S=\text{DM}$. The non-trivial $3\times 3$ rotation block ($R_{ij}^{3\times 3}$) in the middle can be parametrized in terms of three mixing angles $\alpha_1, \alpha_2, \alpha_3$,

\begin{equation}
\begin{pmatrix}
H_2 \\
H_3 \\
H_4
\end{pmatrix}
=
R_{ij}^{3\times 3}(\alpha_1,\alpha_2,\alpha_3)
\begin{pmatrix}
h_2 \\
a_2 \\
h_S
\end{pmatrix}.
\label{rotation}
\end{equation}

\noindent
The non-trivial matrix elements can be written as follows: 
\begin{eqnarray}
M_{11} &=& \lambda_1 v^2 = m_h^2; m_h = 125\text{GeV}, \nonumber \\ 
M_{22} &=& -m_{22}^2 + \left(\frac{\lambda_2'+2\text{Re}[\lambda_5']}{2}\right)v_S^2 + \left(\frac{\lambda_3+\lambda_4+\text{Re}[\lambda_5]}{2}\right)v^2, \nonumber \\
M_{24} &=& vv_S\text{Re}[\lambda_6'+2\lambda_7'], \nonumber\\
M_{33} &=& -m_{22}^2 + \left(\frac{\lambda_2'+2\text{Re}[\lambda_5']}{2}\right)v_S^2 + \left(\frac{\lambda_3+\lambda_4-\text{Re}[\lambda_5]}{2}\right)v^2, \nonumber\\
M_{34} &=& vv_S\text{Im}[\lambda_6' + 2\lambda_7']~\rightarrow\text{Mixing in the scalar sector}, \nonumber\\
M_{44} &=& \frac{1}{6}v_S^2(\text{Re}[\lambda_1''] + 4 \text{Re}[\lambda_2''] + 3 \text{Re}[\lambda_3'']),\nonumber \\
M_{55} &=& -2 \text{Re}[m_S'^2] -\left(\frac{\text{Re}[\lambda_1''] + \text{Re}[\lambda_2'']}{3}\right) v_S^2 - 2 \text{Re}[\lambda_4'] v^2 = m_{\text{DM}}^2. 
\label{matrixelements_alignment}
\end{eqnarray}

\noindent
We mention here that, from now on, we will call the mass parameter $-m_{22}^2$ in Eqs. \ref{matrixelements_alignment} $M^2$.
In the absence of CP-violation, $h_1, h_2, h_S, a_S$ are CP-even scalars and $a_2$ is a CP-odd scalar. One can see from Eq.~\ref{matrixelements_alignment}, 
that a mixing between the CP-odd scalar $a_2$ and CP-even singlet-like scalar $h_S$ is introduced by the $m_{34}$ element, which is proportional to $\text{Im}[\lambda_6' + 2\lambda_7']$. Interesting to note that although $m_S'^2$, $\lambda_1''$, $\lambda_2''$, $\lambda_4'$, $\lambda_5'$, $\lambda_6'$, $\lambda_7'$, $\lambda_8'$, are all in principle complex, only Im$(\lambda_6')$, Im($\lambda_7'$) and Im($\lambda_8'$) can introduce mixing between scalar and pseudoscalars, due to the presence of $\Phi_1^{\dagger} \Phi_2$ term. Notably, hard $Z_2$-breaking of the 2HDM potential ($\lambda_6,\lambda_7 \neq 0$) is essential in this model as well for CP-violation in the scalar potential ({\it i.e.} $\lambda_6',\lambda_7',\lambda_8' \neq 0$). A few comments are in order regarding the roles played by all the complex parameters in Eqs.~\ref{eq:2HDM_potential} and \ref{2hdms}.

The phase of $\lambda_5$ can be rotated away by the redefinition of the fields~\cite{Kanemura:2020ibp}. $\lambda_6=0$, by virtue of the exact alignment condition mentioned earlier in Eqs.~\ref{alignment}. The phase of $\lambda_7$ does not introduce a mixing between the scalar and pseudoscalar states, but it can give rise to CP-violating couplings between scalar and pseudoscalar states. The phases of $\lambda_4'$ and $\lambda_5'$, should be 0 in order for the DM candidate to be stable (see Eq.~\ref{dm}). The phases of $\lambda_1''$ and $\lambda_2''$ do not introduce CP-mixing or CP-violating couplings. 
We list all the complex parameters of the scalar potential and the relevant parameters introducing CP-violation in Table~\ref{cpv_parameters}.

\begin{table}[!h]
\begin{tabular}{|c|c|c|}
\hline
Complex &  CP-mixing and & Only\\
Parameters &  new CP-violating couplings & CP-violating couplings \\
\hline
 $\lambda_5, \lambda_6, \lambda_7$, &  $\lambda_6', \lambda_7', \lambda_8'$ & $\lambda_7$\\
 $\lambda_1''$, $\lambda_2''$, $\lambda_4'$, & & \\
 $\lambda_5'$, $\lambda_6'$, $\lambda_7', \lambda_8'$ & & \\
 \hline
\end{tabular}
\caption{List of all complex parameters: those that can introduce CP-mixing between the scalar and pseudoscalar states and the ones that do not introduce CP-mixing but can give rise to CP-violating trilinear or quadrilinear coupling between the scalar and pseudoscalar states. Please see the discussion on the other couplings in the text.}
\label{cpv_parameters}
\end{table}

We have noticed that the couplings $\lambda_7'$ and $\lambda_8'$ are related via Eq.~\ref{dm}, owing to the DM requirement. Keeping this in mind,
one can see from Table~\ref{cpv_parameters} that, although there are many complex parameters in the scalar potential, in the end, there is one CP-violating phase responsible for a mixing between CP-even and CP-odd states in the scalar mass matrix. 
The CP-violation in the scalar mass matrix can be parametrized in terms of an angle $\theta_{CP} = \frac{\text{Im}[\lambda_6' + 2\lambda_7']}{\text{Re}[\lambda_6' + 2\lambda_7']}$(see Eq.~\ref{matrixelements_alignment}). 

It has been shown in \cite{Kanemura:2020ibp}, that in the Yukawa-aligned 2HDM, in the exact alignment limit, there is no CP-mixing in the mass matrix, and all the mass-eigenstates are CP-eigenstates as well in that case. The major difference between CP-violating 2HDM and its complex-singlet extension is the presence of the additional phase $\theta_{CP}$ in the complex 2HDMS. 

We reiterate that the phase of the coupling $\lambda_7$, namely $\theta_7$, does not appear in the mixing of states. On the other hand, this phase enters the trilinear and quartic scalar-pseudoscalar couplings. Therefore, a non-zero value of $\theta_7$ leads to CP-violating couplings, involving neutral and charged scalars. The presence of this phase is present in both models, namely, CP-violating 2HDM and CP-violating 2HDMS.

\subsection{The Yukawa sector}

The Yukawa Lagrangian of 2HDMS involving $\Phi_1$ and $\Phi_2$, {\it i.e.} the scalar doublets in the Higgs basis is given by
\begin{align}
\mathcal{L}_{\text{yukawa}}=\sum_{k=1}^2 \left( \bar{Q}_L y_{u,k}^{\dagger} \tilde{\Phi}_k u_R + \bar{Q}_L y_{d,k} \Phi_k d_R + \bar{L}_L y_{e,k} \Phi_k e_R  \right). \nonumber
\end{align}

\noindent
As mentioned above, in the absence of $Z_2$ symmetry, in order to avoid tree-level FCNC, Yukawa matrices associated with the two doublets are assumed to be proportional to each other:

\begin{equation}
y_{f,2}=\zeta_f ~y_{f,1},
\end{equation}
where $\zeta$ can be complex and the source of CP-violation. We would like to remind the readers that although such a condition looks ad hoc, it is basically a generalized version of different types of softly broken $Z_2$-symmetric 2HDM. In various types of 2HDM's ({\it e.g.} Type-I, Type-II, Type-X, Type-Y), such a proportionality relation holds only with a specific $\zeta$ involving functions of $\tan\beta$ ($\frac{v_2}{v_1}$).

In terms of the Fermion mass eigenstates, the Yukawa Lagrangian can be written as 

		\begin{eqnarray}
			\mathcal{L}_\text{yukawa}&=&-\sum_{f=u,d,e}\cbra{\bar{f}_L M_f f_R +\sum_{j=1}^3\bar{f}_L \rbra{\frac{M_f}{v} \kappa_f^j} f_R H^0_j+h.c.}\nonumber\\
		&-&\frac{\sqrt{2}}{v}\cbra{-\zeta_u \bar{u}_R (M_u^\dagger V_\text{CKM}) d_L+\zeta_d \bar{u}_L (V_\text{CKM} M_d) d_R+\zeta_e \bar{\nu}_L M_e e_R} H^{+}+h.c.\nonumber\\
        &&
        \label{yukawa}
		\end{eqnarray}
with
		\begin{eqnarray}
\kappa_f^j=\mathcal{R}_{1j}+\sbra{\mathcal{R}_{2j} +i(-2I_f)\mathcal{R}_{3j}} |\zeta_f|e^{i(-2I_f)\theta_f}.
		\end{eqnarray}

\noindent
In 2HDM, in the alignment limit ($R_{ij}=\delta_{ij} $~\cite{Kanemura:2020ibp}), the CP-violation in the Yukawa sector can not come from the CP-mixing in the scalar sector. It must come from the phases of the Yukawa matrices. 
Also, the Yukawa interaction of the 125 GeV Higgs is CP-conserving in the exact alignment limit. 

In CP-violating 2HDMS, the phases $\zeta_f$'s act as an additional source of CP-violation, along with the CP-violating phase ($\theta_{CP})$ pertaining to the mass mixing of the scalar sector and phase of $\lambda_7 (\theta_7)$. 
We reiterate that the crucial difference between CP-violating 2HDM and its complex-singlet extension is that the CP-violation in the scalar sector through the mixing between CP-even and CP-odd states is possible in the latter even in the exact alignment limit. The model files are generated using \texttt{SARAH-v4.14.3}~\cite{Staub:2013tta} and for the particle spectra we used \texttt{SPheno-v4.0.5}~\cite{Porod:2003um}.



\section{Constraints on the model and allowed parameter space}
\label{constraints}
We discuss next the relevant constraints on the model parameter space. CP-violation in our model will be constrained by the electric dipole moment experiments along with other theoretical and experimental constraints. We will discuss their impact in detail. Since we accommodate a dark matter candidate in our model, there are constraints from the observed relic density, the upper bound from the direct detection cross-section and also from the upper bound from the indirect detection experiment. Furthermore, the non-standard scalars in our model must satisfy the limits from direct search at the LEP as well as the LHC.   

\subsection{EDM bounds and allowed parameter space}
The Hamiltonian of the EDM for a non-relativistic particle with the spin ${\vec{S}}$ is as follows:

\begin{equation*}
H_\text{EDM}=-d_f \frac{\vec{S}}{|\vec{S}|}\cdot\vec{E}.
\end{equation*}

\noindent
Under the time reversal transformation: 
$\mathcal{T}(\vec{S})=-\vec{S}$ and $\mathcal{T}(\vec{E})=+\vec{E}$,
the sign of this term $H_\text{EDM}$ is flipped, which implies CP symmetry is broken by virtue of the CPT theorem.

\noindent
In EFT language,
\begin{equation*}
\mathcal{L}_\text{EDM}=-\frac{d_f}{2}\bar{f} \sigma^{\mu\nu}(i\gamma^5)f F_{\mu\nu},
\end{equation*}
\noindent
where $F_{\mu\nu}$ is the electromagnetic field tensor.

The most recent bounds on electron EDM are

$|d_e| < 1.1 \times 10^{-29}\text{e.cm} (ThO) $ ~\cite{Panda:2019dnf},\label{acme}\\
$|d_e| < 4.1 \times 10^{-30}\text{e.cm} (HfF^+) $ ~\cite{Roussy:2022cmp}\label{edmnew}.

In addition to the electric dipole moment of the constituent quarks, the neutron EDM receives the contribution from the chromo-EDM (CEDM) of the quarks $d_q^C$. CEDM of a quark $q$ can be expressed as
		\begin{align}
			\mathcal{L}_\textrm{CEDM}=-\frac{d_q^C}{2}\bar{q} \sigma^{\mu\nu}(i\gamma^5)q G_{\mu\nu}
            \label{sumruleedm}
		,\end{align}
where the $G_{\mu\nu}$ is the QCD field strength tensor.
The nEDM collaboration provides the most updated constraint on the neutron EDM:

~~~~~~~~~~~~~~~~~~~~~~~~~~~~~~~~~~$|d_n|<1.8\times10^{-26}$ e~cm \cite{Abel:2020pzs}.

\noindent
The neutron EDM ($d_n$) can be estimated by using the QCD sum rule\cite{Abe:2013qla}:
		\begin{align}
			d_n=0.79d_d-0.20d_u+e(0.59d_d^C+0.30d_u^C)/g_3
		\label{eq:nEDM}
		,\end{align}

\noindent
where $g_3$ is the QCD SU(3) gauge coupling constant\footnote{The extra factor of $g_3$ in expression~\ref{eq:nEDM} as compared to \cite{Abe:2013qla} is due to an extra factor of $g_3$ in  different definition used for Eq.~\ref{sumruleedm} in \cite{Abe:2013qla}.}.

\begin{figure}[!hptb]
	\centering
        \includegraphics[width=15.0cm,height=4cm]{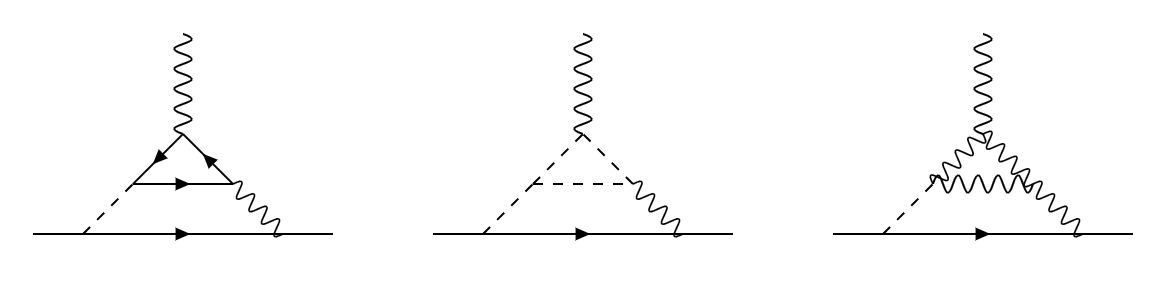}
	\caption{The two-loop Bar-Zee diagrams with fermion (left), scalar (middle) and gauge-boson (right) loop contributions. The solid lines with pointing arrows denote fermion lines (quarks and leptons), the dashed lines denote scalars ($H_{1..4}$ and $H^{\pm}$), and the wavy lines represent gauge bosons ($\gamma, Z, W$). }
	\label{barzee}
\end{figure}

\noindent
The major contribution to the EDM and CEDM comes from the two-loop Bar-Zee diagrams shown in Fig.~\ref{barzee}. The contribution to the Bar-Zee diagrams can be decomposed into fermion, gauge boson and scalar running in the loop: 

		\begin{equation*}
			d_f=d_f(\textrm{fermion})+d_f(\textrm{Higgs})+d_f(\textrm{gauge}).
		\end{equation*}

\noindent
Each contribution $d_f(X)$ further consists of contributions from $\gamma,Z$ and $W$ in an internal leg, 
		\begin{equation*}
			d_f(X)=d_f^\gamma(X)+d_f^Z(X)+d_f^W(X).
            \label{edmeq}
		\end{equation*}

\noindent
The detailed expressions for the aforementioned loop contributions are given in Appendix~\ref{BarZeeformulae}. 
The gauge boson loops contribute negligibly in the alignment limit, since the couplings of the gauge boson to non-standard scalars are abysmal in the alignment limit (since $R_{1j}\approx 0$ in Eqs.~\ref{gauge1} and \ref{gauge2} in the alignment limit). On the other hand, the fermion and scalar boson loop can contribute at equivalent strength. Therefore, it is crucial to take into account both contributions. One-loop contributions are suppressed by at least 4-5 orders of magnitude compared to the two-loop diagrams. 
First, we will study the impact of electron EDM on our parameter space. 

In our model, there are three sources of CP-violation. 
\begin{itemize}
\item CP-mixing between neutral scalar and pseudoscalar eigen-states, the relevant phase in this case is $\theta_{CP}$.
\item The phase $\theta_7$ of $\lambda_7$ coupling, which does not appear in the CP-mixing in the neutral scalar sector, but introduces CP-violating trilinear and quadrilinear vertices between scalar and pseudoscalar states. 
\item The phases in the Yukawa matrix, namely $\theta_u, \theta_d$ and $\theta_e$, phases associated with $\zeta_u, \zeta_d$ and $\zeta_e$ respectively. 
\end{itemize}

\begin{figure}[!hptb]
	\centering
\includegraphics[width=17.0cm,height=7cm]{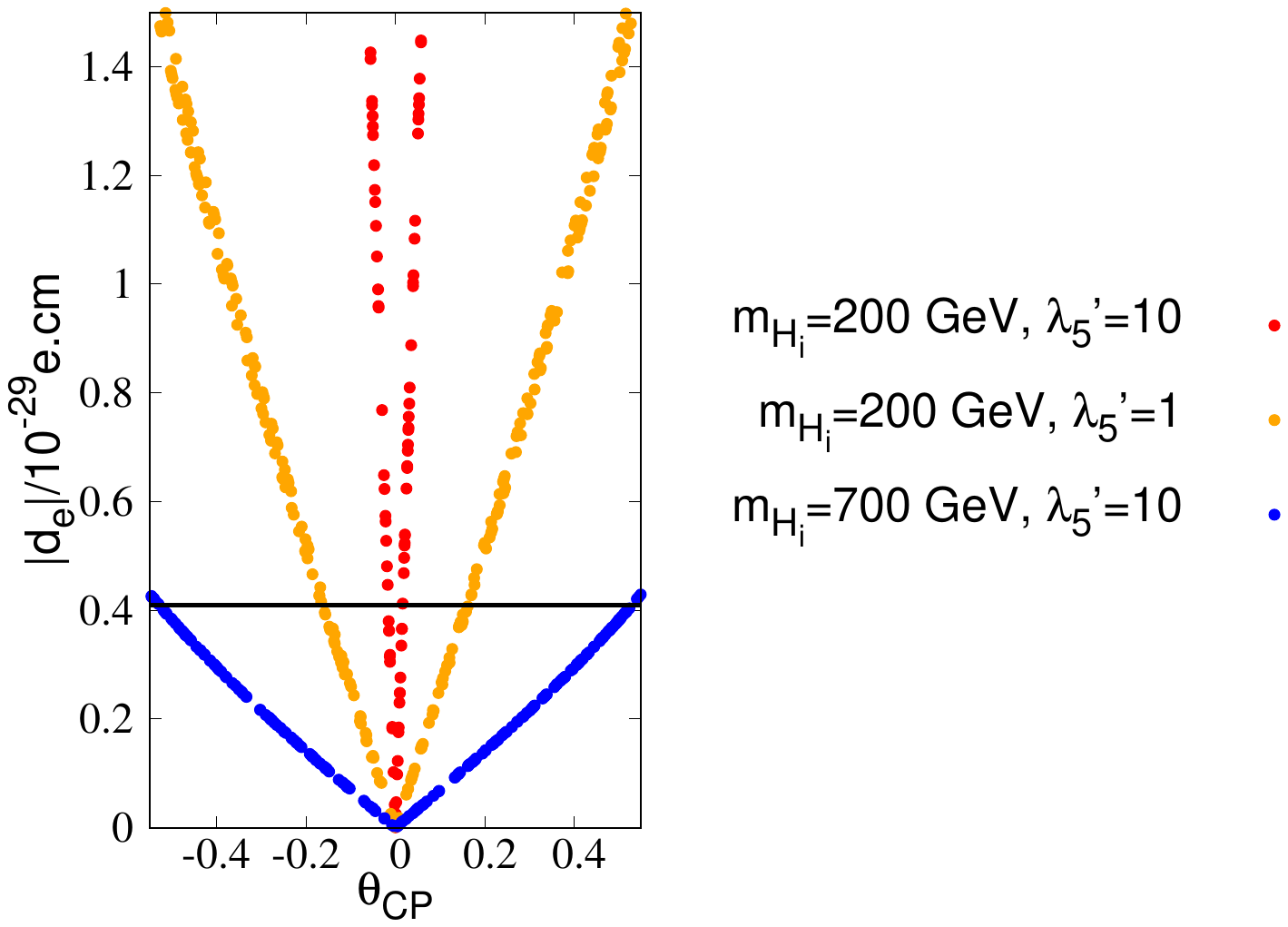}
	\caption{Constraints on $\theta_{CP}$ from the most recent electron EDM bound with $\theta_7=0, \theta_f=0$, $|\lambda_7|=0.3, \lambda_2'=0$.}
	\label{thetacp}
\end{figure}

\noindent
We first show the impact of EDM constraints on $\theta_{CP}$ in Fig.~\ref{thetacp}. As we mentioned, contribution to EDM comes from both fermion and scalar loops. The size of the scalar loop contribution increases with $\lambda_2'+2\lambda_5'$ (see Eq.~\ref{definitions} in Appendix~\ref{BarZeeformulae}). We present the results for two values of $\lambda_5'$, with $\lambda_2'=0$. Furthermore, amplitude decreases with increasing scalar masses. In order to demonstrate that effect, we choose two mass values for the non-standard scalars $H_i$'s. 
One can see from Fig.~\ref{thetacp}, the limits on $\theta_{CP}$, the phase responsible for CP-mixing among neutral scalars from electron EDM, when the other two sources of CP-violation are absent {\it i.e.} $\theta_7=\theta_f=0$. The limits are more stringent when the scalar masses are low, and $\lambda_5'$ is large, due to a large contribution. 

\begin{figure}[!hptb]
	\centering
        \includegraphics[width=7.5cm,height=6.0cm]{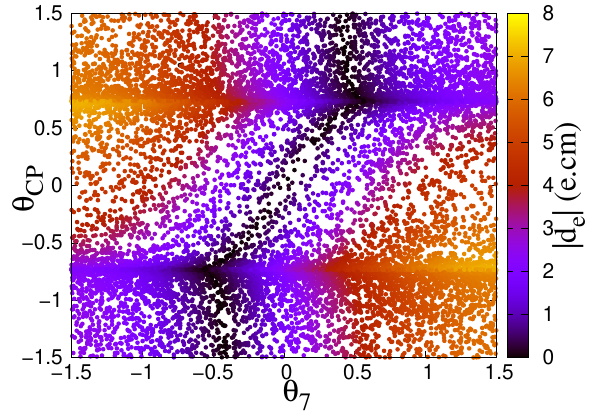}
	\caption{Constraints on $\theta_{CP}$ and $\theta_7$ from the most recent electron EDM bound. $\theta_f=0$, $\lambda_7=0.3, \lambda_2'=0, \lambda_{5}'=1$,
   Masses of non-standard scalars  $m_{H_i}\approx 200$ GeV. The electron EDM $|d_e|$ for the parameter points in units of $10^{-29}$ is shown in the color axis. }
	\label{thetacp_theta7}
\end{figure}

Next, we consider the impact of two phases, namely $\theta_{CP}$ and $\theta_7$ simultaneously, while $\theta_f=0$. The two contributions from the fermion and scalar loop appear with opposite sign, and therefore, there is a possibility of cancellations between them. We show in Fig.~\ref{thetacp_theta7} the possibility of such cancellation and the resulting allowed region. One can see in Fig.~\ref{thetacp_theta7} a narrow strip in black where the EDM bound will be satisfied.


We move further to explore the impact of the third source of CP-violation in our model. In \cite{Kanemura:2020ibp}, the CP-violating Yukawa-aligned 2HDM is considered. In that model, in the alignment limit, the only sources of CP-violation are the phase of $\lambda_7$ ($\theta_7$) and phases in the Yukawa matrices ($\theta_f$). In the complex singlet extension of that model, which we consider in this work, we have another source of CP-violation $\theta_{CP}$. It becomes a natural question: how does our scenario compare with \cite{Kanemura:2020ibp}? We demonstrate this next.

\begin{figure}[!hptb]
	\centering
        \includegraphics[width=6.0cm,height=5.0cm]{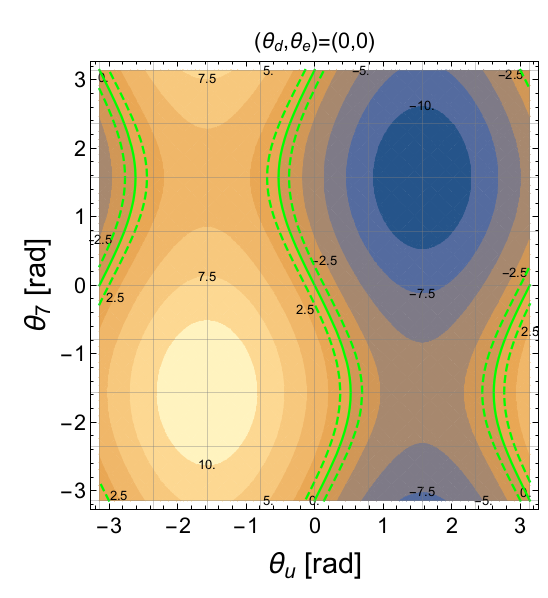}
        \includegraphics[width=6.0cm,height=5.0cm]{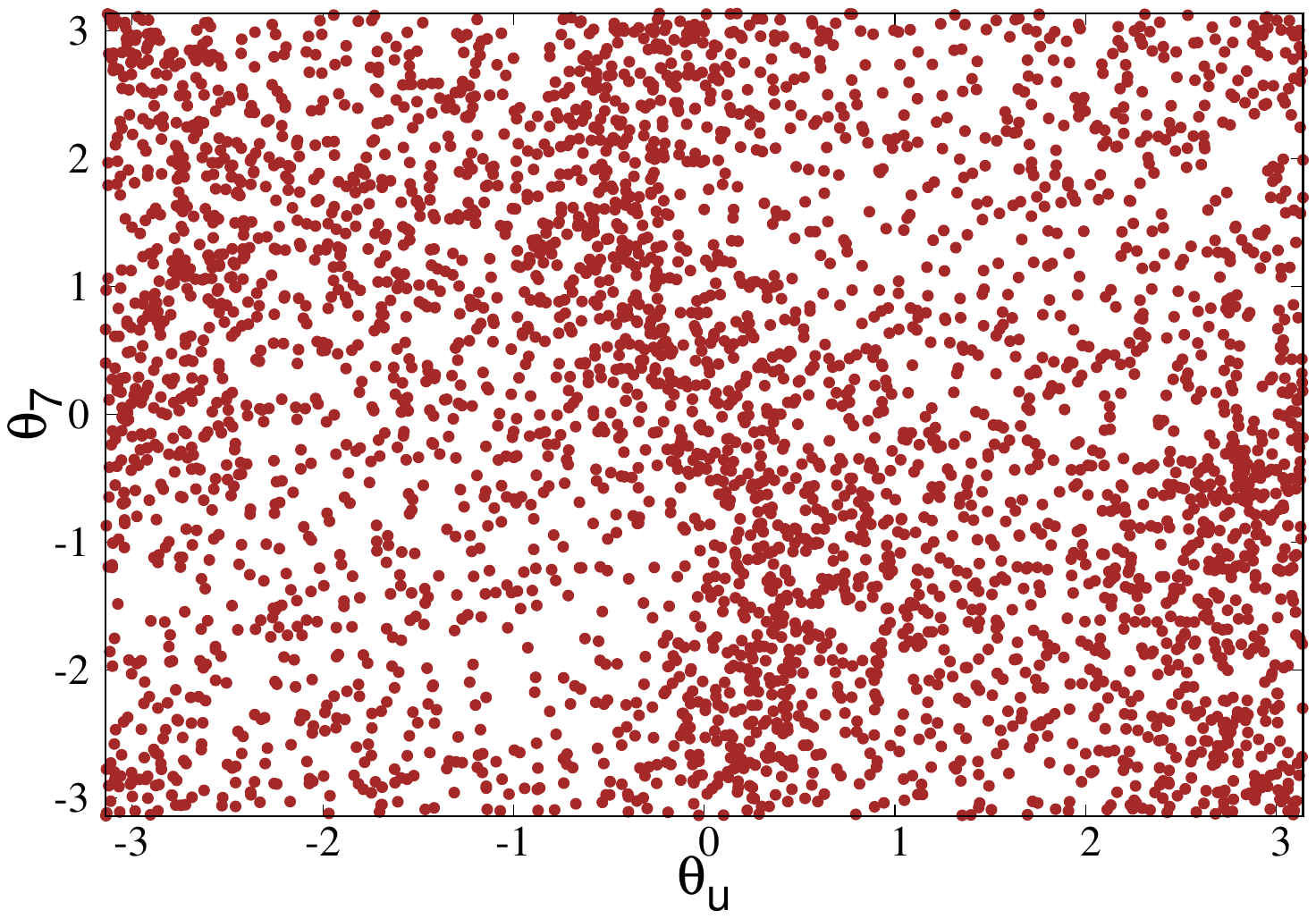} \\
	\caption{(left) Electron-EDM allowed region in $\theta_u-\theta_7$ parameter space in Yukawa-aligned 2HDM with rest of the parameters given in Table~\ref{2hdminput} (Figure borrowed from \cite{Kanemura:2020ibp}). (right) In 2HDMS, electron-EDM allowed region in $\theta_u-\theta_7$ parameter space with inputs from Table~\ref{2hdminput} and the additional relevant parameters are varied as follows: $-\pi < \theta_{CP} < \pi$ and $0 < \lambda_5' < 10$.}
	\label{kanemura_compare}
\end{figure}

\noindent
We borrow Fig.~\ref{kanemura_compare}(left) from \cite{Kanemura:2020ibp} for comparison. This plot is in the context of Yukawa-aligned 2HDM. One can see in Fig.~\ref{kanemura_compare}(left), that a cancellation along the region with green boundary in the $\theta_7-\theta_u$ plane makes it possible to evade the EDM bounds. We study next the same parameter space shown in Fig.~\ref{kanemura_compare}(left), in the context of 2HDMS. When the additional phase $\theta_{CP}$ is varied between [$-\pi,\pi$], we see that the entire region in the $\theta_u-\theta_7$ plane becomes allowed by EDM constraints. We mention that for this comparison we use the older EDM bound Eq.~\ref{acme}, instead of the most recent EDM bound given in Eq.~\ref{edmnew}, because the older bound was imposed on 2HDM parameter space in \cite{Kanemura:2020ibp}. We varied another relevant parameter $0 < |\lambda_5'| < 10$, while $\lambda_2'$ is chosen to be 0. We use the same values for the 2HDM parameters as in \cite{Kanemura:2020ibp}, shown in Table~\ref{2hdminput}.

\begin{table}
\centering
\begin{tabular}{|c|c|c|c|c|c|c|c|c|c|c|}
\hline
$m_{H_1}$ & $m_{H_2}$ & $m_{H_3}=m_{H^\pm}$ & $M$ & $|\zeta_u|$ & $|\zeta_d|$ & $|\zeta_e|$ & $|\lambda_7|$ & $\lambda_2$ & $\theta_{d,e}$ \\
in GeV &in GeV & in GeV & in GeV & & & & & & in radian \\
\hline
125 & 280 & 230 & 240 & 0.01 & 0.1 & 0.5 & 0.3 & 0.5 & 0 \\
\hline
\end{tabular}
\caption{Input values for Yukawa-aligned 2HDM scenario corresponding to Fig.~\ref{kanemura_compare}(left).}
\label{2hdminput}
\end{table}

\begin{figure}[!hptb]
       	\centering
        \includegraphics[width=7.5cm,height=6.0cm]{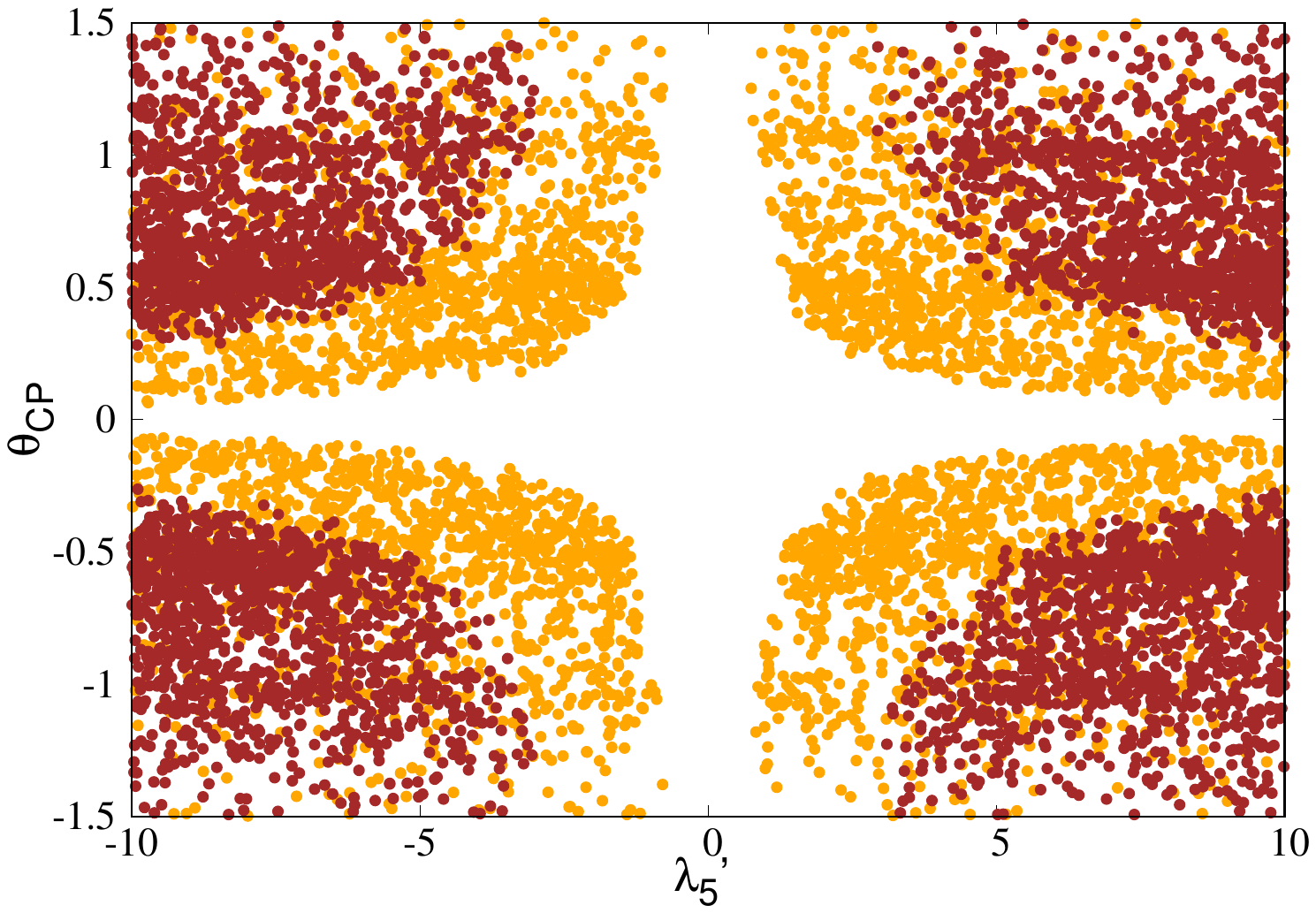}
       
	\caption{The allowed region in CP-violating 2HDMS with $\sbra{\theta_u,\theta_7}=\sbra{\frac{\pi}{2},\frac{\pi}{2}}$, all other parameters kept fixed as in Table~\ref{2hdminput}. The orange points correspond to $m_{H_4} = 200$ GeV, and the maroon points correspond to $m_{H_4}=700$ GeV.}
     \label{thetacpl5p}
\end{figure}

\begin{figure}[!hptb]
       	\centering
        \includegraphics[width=7.5cm,height=6.0cm]{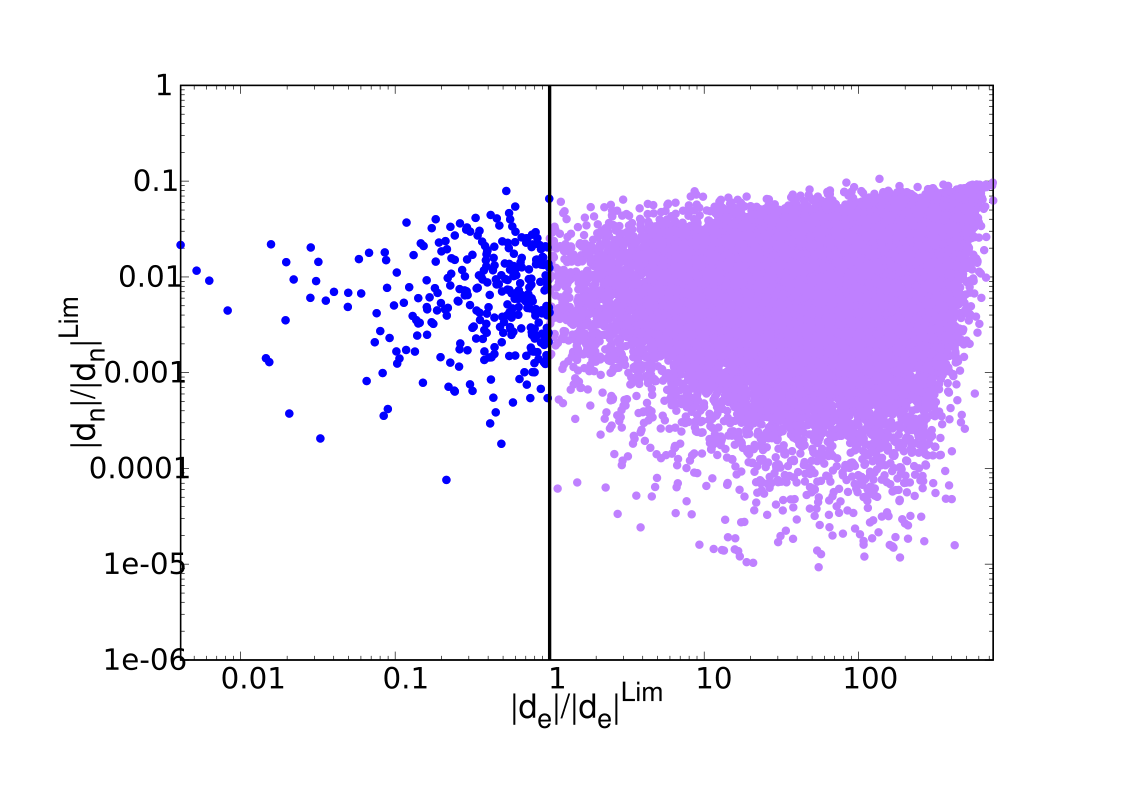}
       
	\caption{The ratio of the electron EDM and observed upper limit on electron EDM $\frac{|d_e|}{|d_e|^{Lim}}$ on the x axis and the ratio of the neutron EDM and observed upper limit on neutron EDM $\frac{|d_n|}{|d_n|^{Lim}}$ on the y axis.  Purple points are excluded by the electron EDM constraints, and blue points are allowed by the electron EDM constraints. }
     \label{de_dn}
\end{figure}

Next we chose a point from Fig.~\ref{kanemura_compare}(left), $\sbra{\theta_u,\theta_7}=\sbra{\frac{\pi}{2},\frac{\pi}{2}}$ which in CP-violating 2HDM leads to $d_e = -12.7\times 10^{-29}$ e.cm. Therefore, such a parameter point is clearly ruled out by EDM bounds in CP-violating 2HDM. Now, we study this particular parameter point in the context of CP-violating 2HDMS while varying the additional phase $\theta_{CP}$ and $\lambda_5'$, and we show the allowed parameter space in Fig.~\ref{thetacpl5p}.
From Fig.~\ref{thetacpl5p}, it is clear that the presence of a third source of CP-violation and extra degrees of freedom alleviates the fine-tuned cancellation of CP-violating 2HDM. A larger parameter space is allowed by EDM experiments in CP-violating 2HDMS. Also, a lighter non-standard scalar mass ($m_{H_4}$) and a larger $\lambda_5'$ value ensure a more effective cancellation between various contributions and a larger allowed parameter space.

Apart from electron EDM, we also investigated the neutron EDM in our model and its impact on the parameter space. As we discussed earlier, the neutron EDM receives contributions from quark EDM as well CEDM's. There are additional contributions to the neutron EDM from the Weinberg operator $\frac{1}{3}C_WG^a_{\mu\nu}\tilde{G}^b{}^{\nu\sigma}G^c{}_{\sigma}{}^{\mu}$ and the four-Fermi interaction $C_{ff'}(\bar{f}f)(\bar{f}'i\gamma_5f')$.
However, it was shown in earlier works ~\cite{Bernreuther:1990jx,Cheung:2014oaa,Jung:2013hka} the contribution from $C_W, C_{ff'}$ are sub-leading. Therefore, we use the EDM and CEDM contribution of quarks via the sum-rule (Eq.~\ref{eq:nEDM}) for our calculation. We have computed the neutron EDM for our parameter space with an extensive scan (Table~\ref{scan_relic}). We show a comparison between the electron EDM and neutron EDM for the scanned points in Fig.~\ref{de_dn}. It is clear from the figure that the limits from the non-observation of electron EDM is much stronger than those of neutron EDM for the scanned points of our model. Therefore, from now on, EDM constraints will imply those of the electron EDM.




\subsection{Constraints from and interplay with dark matter sector}

Now we examine the impact of the CP-violating phases on dark matter phenomenology since we are interested in the model 2HDMS with complex parameters, where CP-violation and dark matter can be accommodated simultaneously. In order to understand the role of the phases in the DM phenomenology, we list the trilinear and quartic couplings that take part in annihilation and therefore observed relic density as well as direct and indirect detection cross-section:

{\small{\begin{eqnarray*}
\lambda_{a_S a_S h_1 h_1} &=& -\lambda_4', \nonumber \\
\lambda_{a_S a_S h_1 h_2} &=& \frac{1}{2}(\text{Re}[\lambda_6'] - 2\text{Re}[\lambda_7']), \nonumber \\
\lambda_{a_S a_S h_2 h_2} &=& \frac{1}{4} (\lambda_2' - 2\lambda_5'), \nonumber \\
\lambda_{a_S a_S h_1 a_2} &=& -\frac{1}{2}(\text{Im}[\lambda_6'] - 2\text{Im}[\lambda_7']), \nonumber \\
\lambda_{a_S a_S a_2 a_2} &=& \frac{1}{4} (\lambda_2' - 2\lambda_5'), \nonumber \\
\lambda_{a_S a_S h_S h_S} &=& -\frac{1}{8} (\lambda_1'' - \lambda_3''), \nonumber \\
\lambda_{a_S a_S h_1} &=& -2 v \lambda_4', \nonumber \\
\lambda_{a_S a_S h_2} &=& \frac{1}{2} v (\text{Re}[\lambda_6'] - 2\text{Re}[\lambda_7']), \nonumber \\
\lambda_{a_S a_S h_S} &=& -\frac{1}{4} v_S (\lambda_1'' - \lambda_3''), \nonumber \\
\lambda_{a_S a_S a_2} &=& -\frac{1}{2} v (\text{Im}[\lambda_6'] - 2\text{Im}[\lambda_7']). 
\label{dm_couplings}
\end{eqnarray*}}}

\noindent
One should note that the phases of $\lambda_6'$ and $\lambda_7'$ appear in the dark portal couplings. However, their combination here is different compared to the CP-mixing in the neutral scalar sector.

\begin{figure}[!hptb]
	\centering
        \includegraphics[width=10.0cm,height=4.7cm]{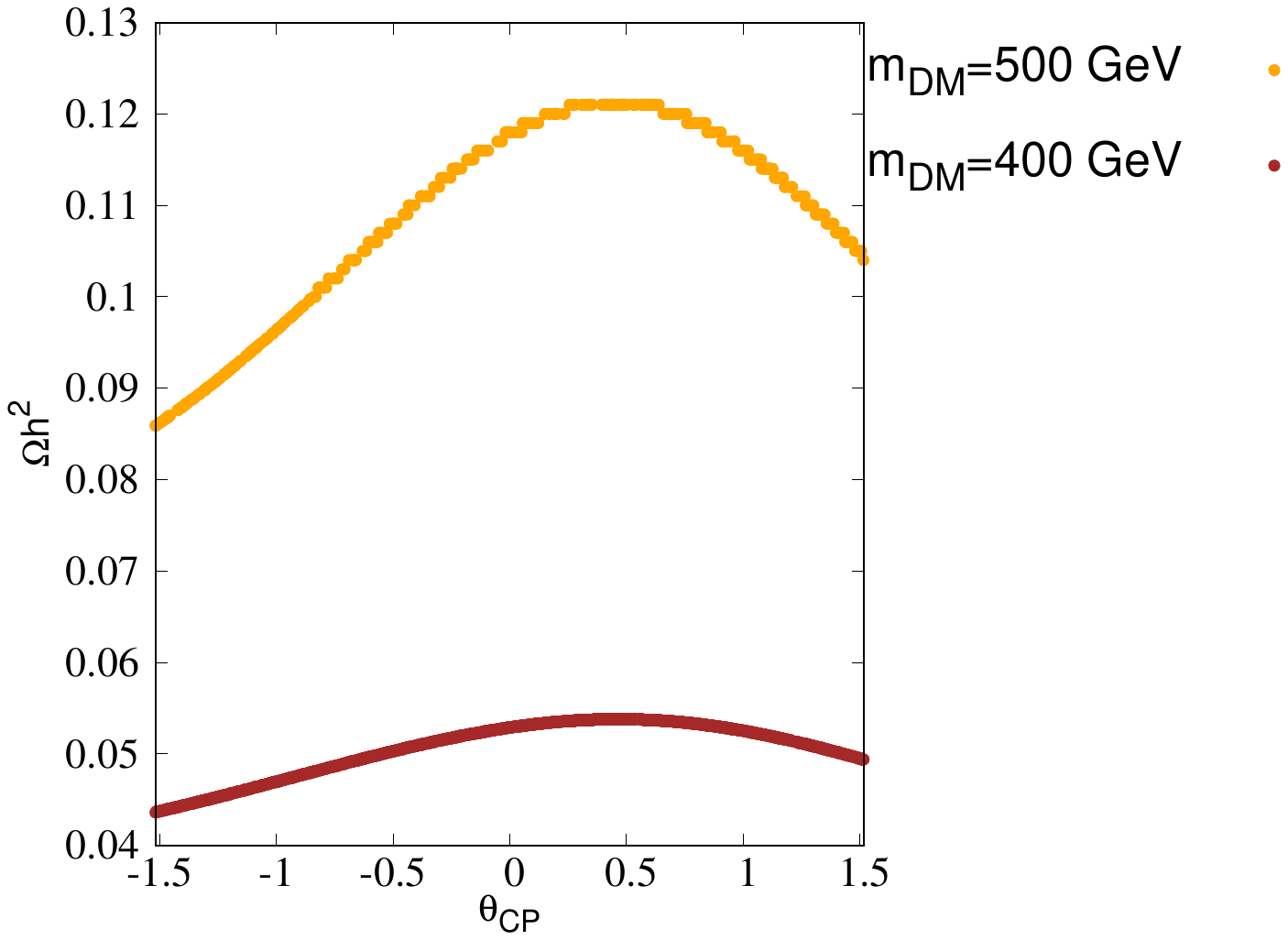}
	\caption{The variation of relic density as a function of the CP-violating phase $\theta_{CP}$, when all other parameters are kept fixed ($\lambda_2' = 0.1, \lambda_5'=0.052, |\lambda_6'+2\lambda_7'|=0.1, \lambda_1''=0.2, \lambda_3''=0.1, \lambda_4'=0.01, \lambda_7=0.3, \tan\beta=1.5, v_S= 1\text{TeV},\theta_f=\theta_7=0, m_{H_i}=200$ GeV). }
	\label{relic_thetacp}
\end{figure}

\begin{figure}[!hptb]
	\centering
        \includegraphics[width=7.0cm,height=5.7cm]{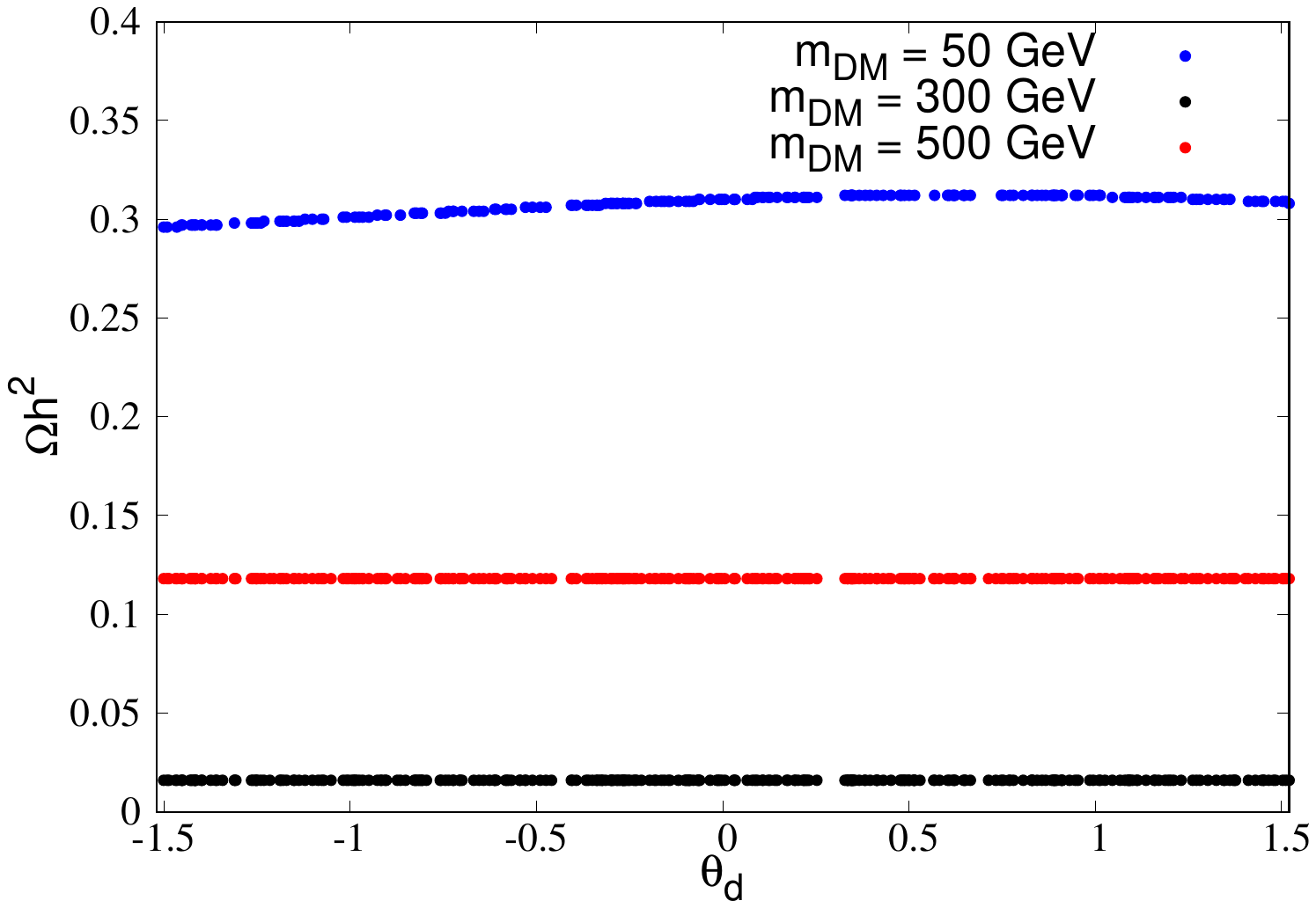}
	\caption{ The variation of relic density as a function of the CP-violating phase $\theta_{d}$, when all other parameters are kept fixed ($\lambda_2' = 0.1, \lambda_5'=0.052, |\lambda_6'+2\lambda_7'|=0.1, \lambda_1''=0.2, \lambda_3''=0.1, \lambda_4'=0.01, \lambda_7=0.3, \tan\beta=1.5, v_S= 1\text{TeV}, \theta_{\text{CP}}=\theta_u=\theta_e=\theta_7=0, m_{H_i}=200$ GeV). }
	\label{relic_thetad}
\end{figure}

In Fig.~\ref{relic_thetacp}, we show scenarios where the portal couplings are chosen in such a way that direct detection and indirect detection bounds are satisfied and at the same time, we are close to the observed relic density $\Omega h^2 \approx 0.12$. We kept all other parameters fixed and varied the CP-violating phase $\theta_{CP}$, in order to show the effect of it on the observed relic density. We have shown our results for two DM masses. One can see that variation of $\theta_{CP}$ can induce around 50\% variation in the observed relic density. 

In Fig.~\ref{relic_thetad}, we show the variation of relic density as a function of the phase of the Yukawa matrices $\theta_d$. We can see the relic density is almost independent of the phase of the Yukawa matrices. This is due to the fact that the major annihilation channels for DM pairs are into the scalar states when kinematically feasible. Only for light dark matter, when the scalar and gauge boson modes of annihilation are kinematically disfavored, the major channel becomes DM DM $\rightarrow b\bar b$. That is when we can observe a small but non-trivial dependence of the relic density on the phase $\theta_d$ for $m_{\text{DM}}=50$ GeV. We have checked that there is practically no dependence of the relic density on the remaining phases, $\theta_u, \theta_e$ and $\theta_7$.

\begin{figure}[!hptb]
	\centering
        \includegraphics[width=7.0cm,height=5.7cm]{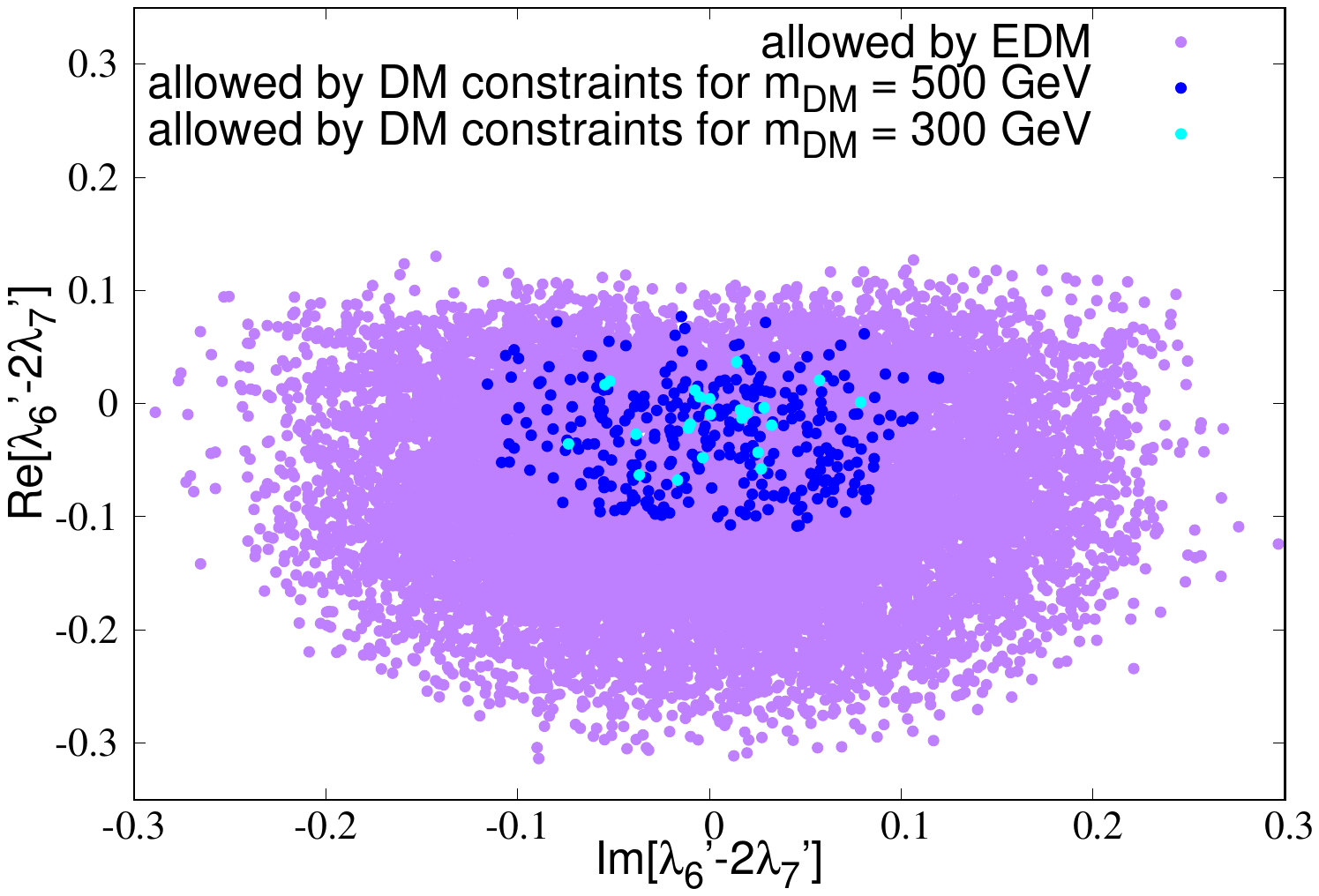}
       \includegraphics[width=7.0cm,height=5.7cm]{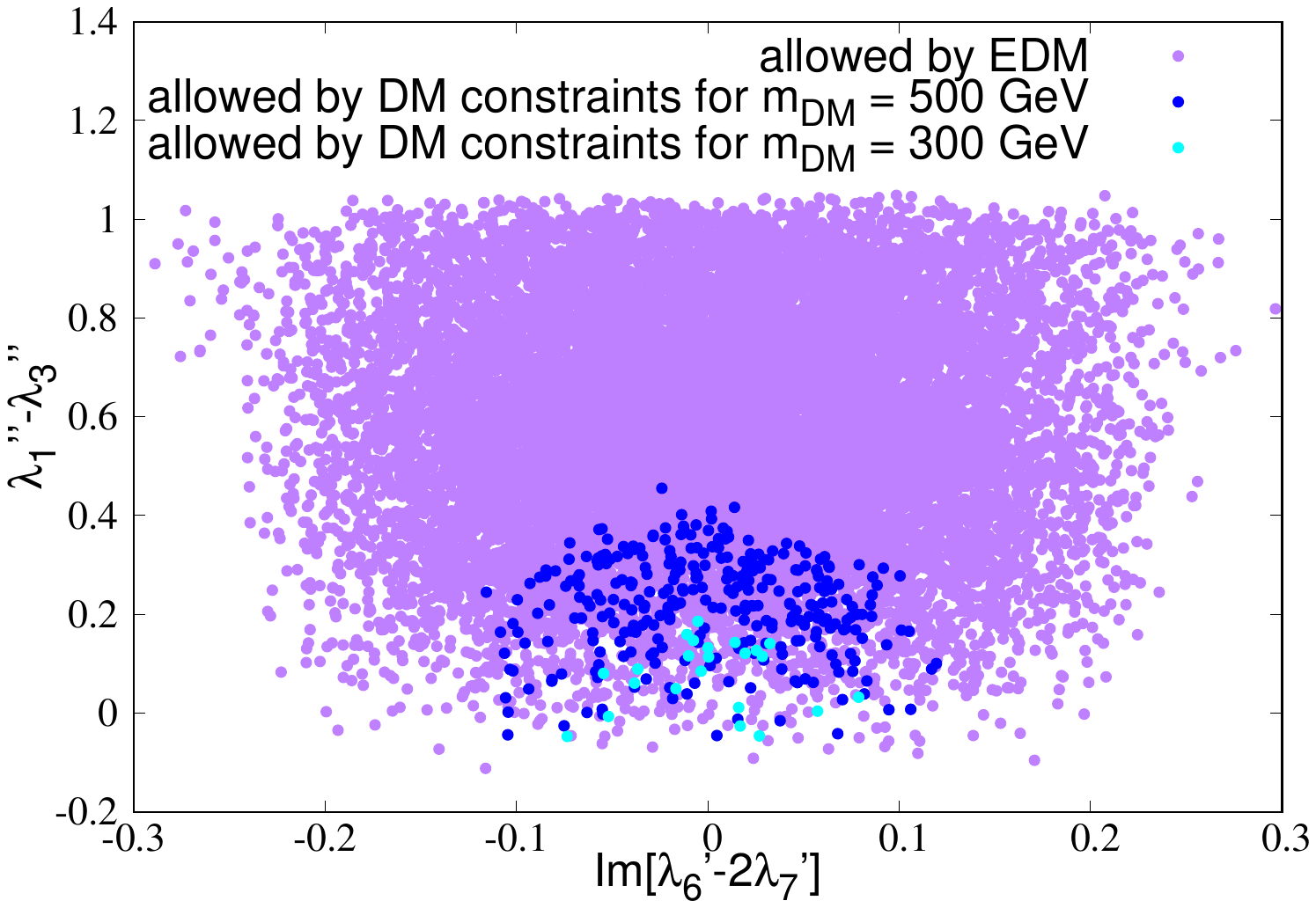} \\
            \includegraphics[width=7.0cm,height=5.7cm]{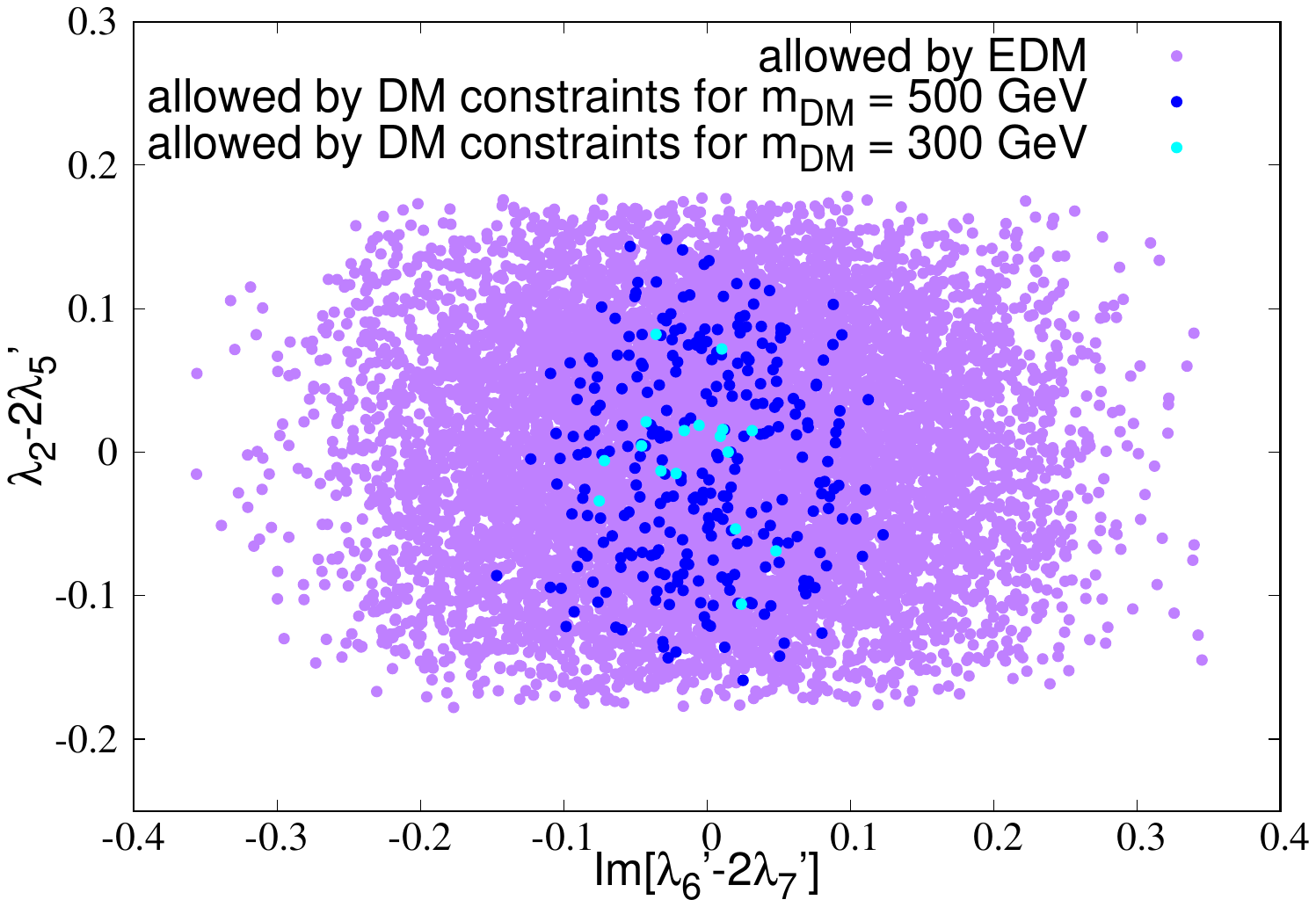}  
            \includegraphics[width=7.0cm,height=5.7cm]{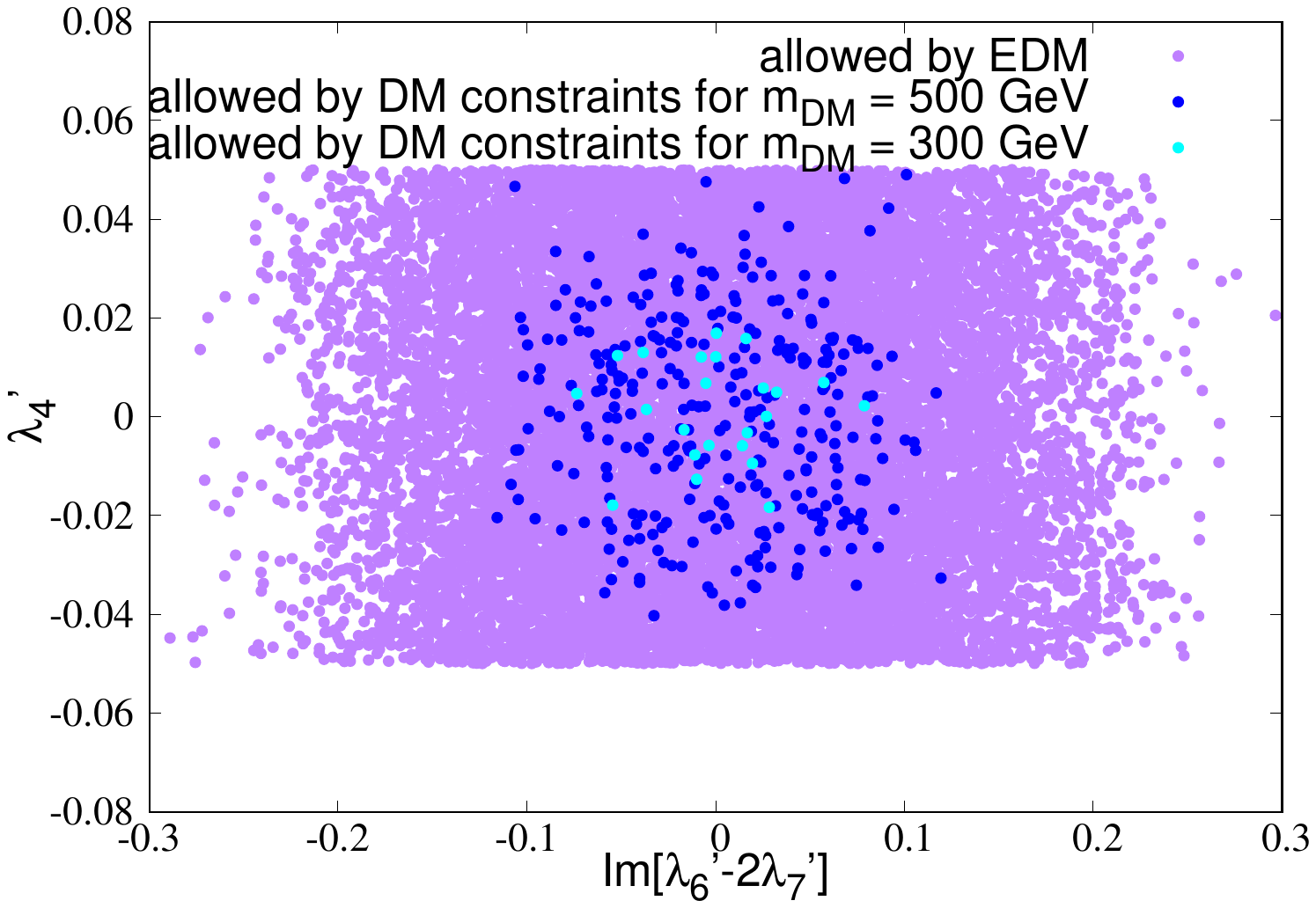}
	\caption{Allowed parameter space from EDM as well as dark matter constraints {\it e.g.} direct detection, indirect detection, and observed relic density, for two different dark matter masses.}
	\label{relic_scan}
\end{figure}

After examining the role of different CP-violating phases in the relic density calculation, we do a parameter scan (scan range given in Table~\ref{scan_relic}) and identify regions that satisfy the observed relic density $\Omega h^2 = 0.1191 \pm 0.0010$ from Planck~\cite{Planck:2018vyg}. We also impose upper bounds from direct detection experiments, namely LUX-ZEPLIN~\cite{LZ:2022lsv}, and indirect detection experiments, namely Fermi-LAT~\cite{Fermi-LAT:2011vow,Fermi-LAT:2016uux}. The coupling combinations that enter the trilinear and quadri-linear couplings between DM and scalar states, and therefore crucial in the calculation of the aforementioned DM observables, are listed in Eqs.~\ref{dm_couplings}.
These coupling combinations get direct constraints from the dark matter sector, and therefore, the impact of such constraints on these parameters is worth studying. The predictions of the model for the direct and indirect detection cross sections and relic abundance are calculated using~\texttt{micrOmegas-v5.2.13}~\cite{Belanger:2010pz}.

\begin{table}
\centering
{\small{\begin{tabular}{ccc}
\hline
200 GeV $< m_{H_i} \approx m_{H^\pm} <$ 400 GeV, &	200 GeV $< M <$ 400 GeV,
&	$-\pi < \alpha_{1,2,3} < \pi $,
\\ \hline
0$<|\lambda_5'|,|\lambda_2'|,|\lambda_7'|,|\lambda_2''|,
|\lambda_4'|<0.1$, & $-\pi < \theta_7' < \pi$ & 100 GeV $< v_S < 1000$ GeV,
\\ \hline
0$<|\lambda_3''| <$ 1 & 0$<|\lambda_7| <$ 1 & -$\pi < \theta_7 < \pi$\\
\hline
$-\pi < \theta_{u,d,e} < \pi$ & $\epsilon_u=0.01, \epsilon_d=0.1, \epsilon_e=0.5$ & $\tan\beta=1.5$\\
\hline
\end{tabular}}}
\caption{Scan ranges for relevant input parameters. The rest of the couplings can be calculated from the input parameters using the mass relations. }
\label{scan_relic}
\end{table}

We show in Fig.~\ref{relic_scan}, the allowed region in the parameter space spanned by various coupling combinations. In all the plots, the purple points satisfy EDM bounds, the blue points satisfy dark matter constraints (observed relic density, upper bound from direct and indirect detection experiments) for DM mass=500 GeV, and the cyan points satisfy the same constraints for $m_{\text{DM}}=300$ GeV.

\subsection{Constraints from HiggsBounds, perturbative unitarity and B-physics}

Next, we focus on the theoretical and experimental constraints on our model parameter space. The constraints we considered are as follows.

\begin{itemize}
    \item \textbf{Tree-level unitarity:} 
    The requirement of the model to be unitary at tree-level constrains the eigenvalues of the scattering matrices between the scalars and the longitudinal components of the gauge bosons to be lower than $\frac{1}{2}$~\cite{Goodsell:2018tti}. This condition is checked for each point of the scans using \texttt{SPheno-v4.0.5}~\cite{Porod:2003um}.
\item \textbf{B-physics:} Constraints from B-physics come from the following bounds: 
    \begin{align*}
        BR(b \rightarrow s \gamma) 
        &= (3.55\pm0.24\pm0.09)\times 10^{-4}~\text{\cite{BaBar:2012fqh}},\\
        BR(B_s \rightarrow \mu^+\mu^-)
        &=(3.2^{+1.4 +0.5}_{-1.2 -0.3})\times10^{-9}~\text{\cite{LHCb:2013vgu,CMS:2013dcn}}.
    \end{align*}
\item \textbf{Oblique parameters:} The electroweak precision tests constrain the $STU$ parameters as follows.~\cite{ParticleDataGroup:2020ssz}:
    \begin{align*}
        S &= 0.02 \pm 0.1, \\
        T &= 0.07\pm0.12, \\
        U &= 0.00\pm0.09. 
    \end{align*}

     The predictions for the STU parameters for our parameter space are obtained from~\cite{Grimus:2007if,Grimus:2008nb}. 
\item \textbf{HiggsBounds and Higgssignal:} The constraints from LEP~\cite{ALEPH:2013htx}, ATLAS~\cite{higgssumatlas} and CMS~\cite{higgssumcms} on the heavy Higgs searches and the $125 \, \text{GeV}$ Higgs signal strength measurements~\cite{ATLAS:2020qdt} are taken into account.
\end{itemize}

\begin{figure}[!hptb]
	\centering
        \includegraphics[width=7.0cm,height=5.7cm]{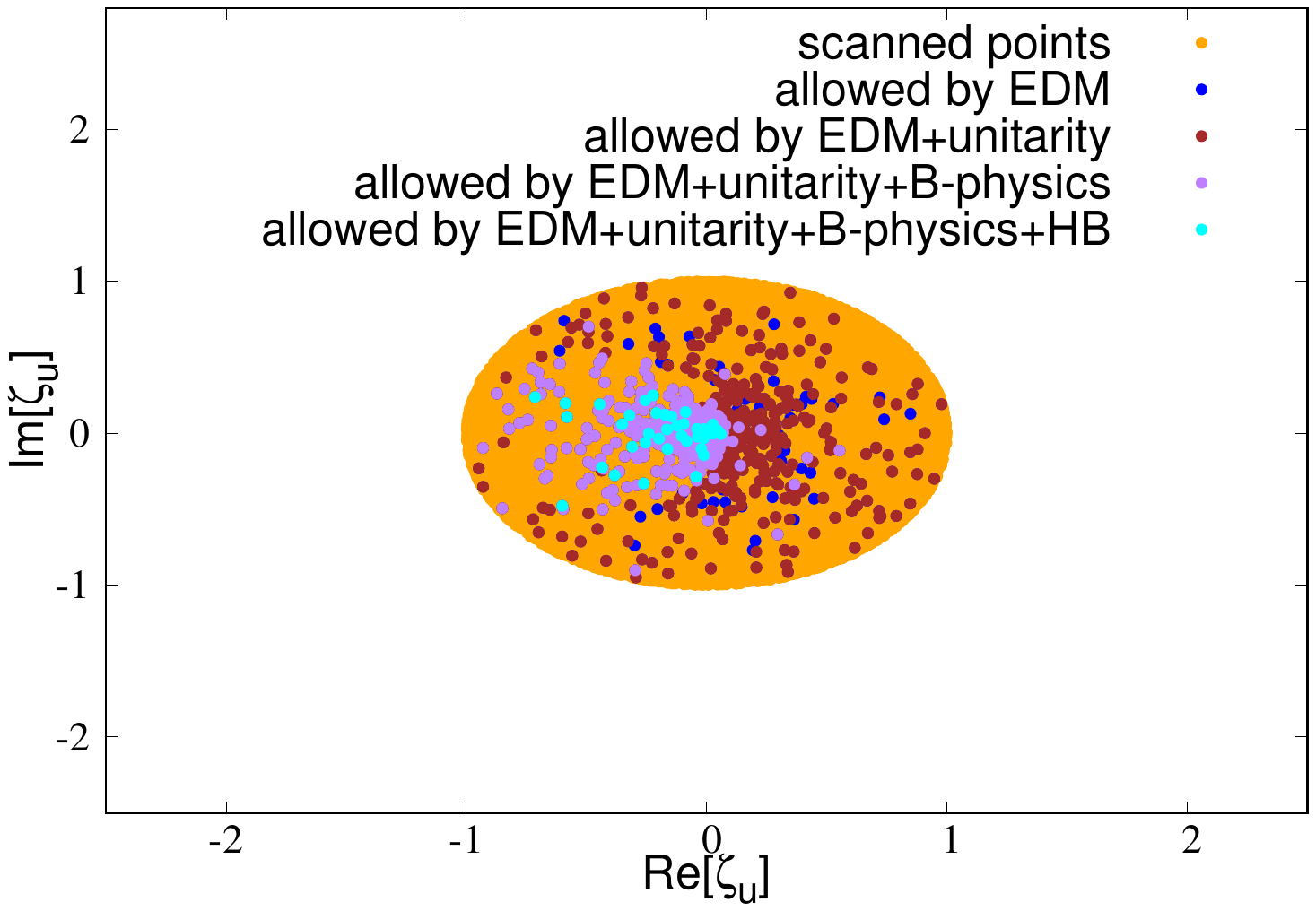}
       \includegraphics[width=7.0cm,height=5.7cm]{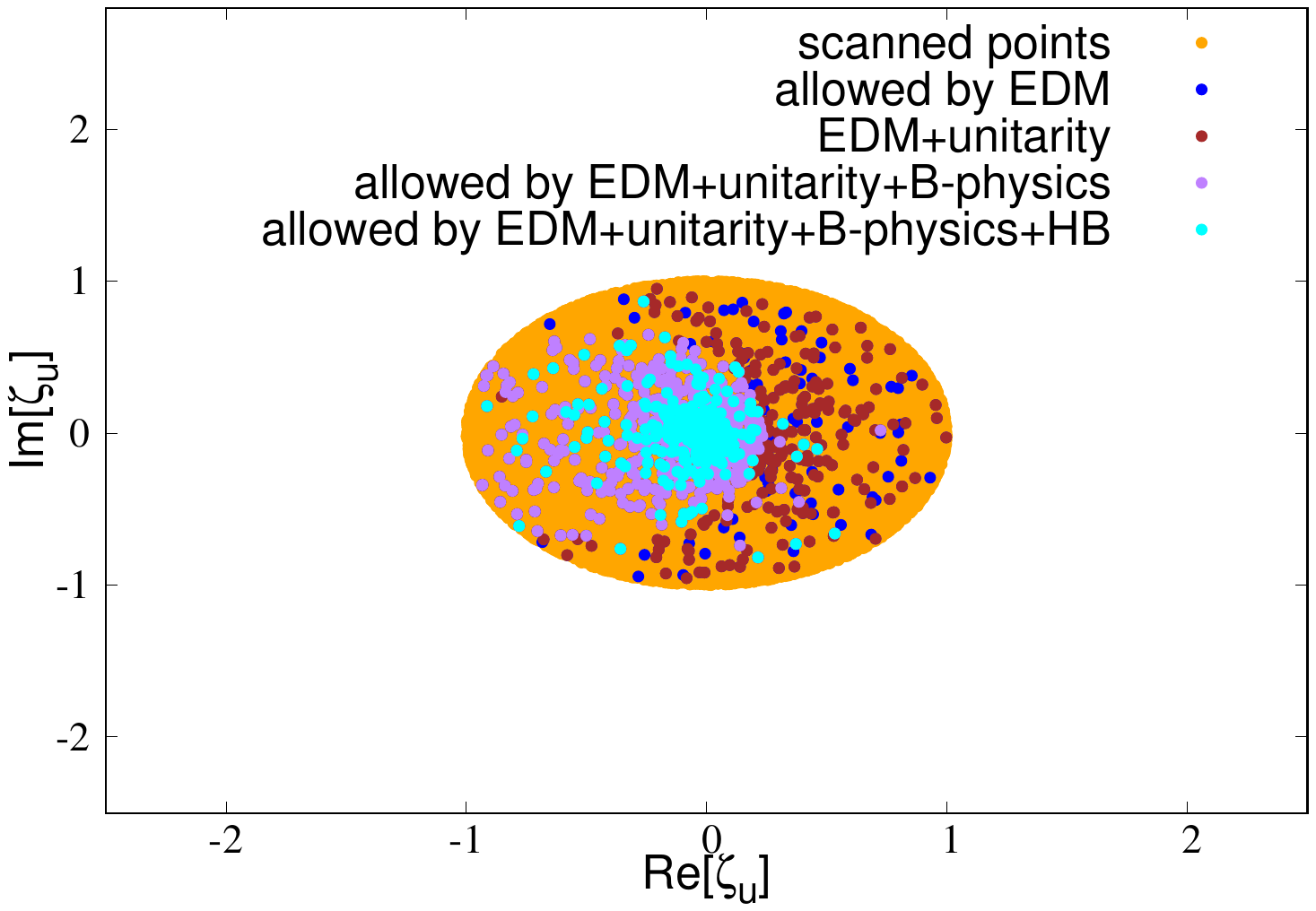} \\
            
	\caption{ The allowed parameter space from EDM as well as unitarity, B-physics and HiggsBounds. The non-standard scalar masses are at $\sim$ 200 GeV (left) and $\sim$ 400 GeV (right). The rest of the scan range is the same as in Table~\ref{scan_relic}. }
	\label{edm_bphyics_unitarity_HB}
\end{figure}

\noindent
We have seen earlier that the low scalar mass region looks interesting from the EDM point of view, since low scalar masses enable effective cancellation between the top loop and scalar loop Bar-Zee diagrams. Also, low-mass non-standard scalars are interesting from the point of view of early detection at future colliders. However, it is well-known that the low mass non-standard scalars of 2HDM's are strongly constrained by the direct search of BSM scalars implemented in HiggsBounds~\cite{Bechtle:2008jh,Bechtle:2011sb,Bechtle:2013wla,Bechtle:2020pkv} as well as limits coming from B-physics observables. This is true especially in Type-II 2HDM. On the other hand, Type-I or Type-X 2HDM are comparatively relaxed from B-physics considerations~\cite{Enomoto:2015wbn}. Interestingly, in our model, we have the free parameters $\zeta$'s that give us more freedom in terms of Yukawa couplings, and this opens up allowed parameter space allowed by HiggsBounds, B-physics, even in the low-mass region, which is also consistent with EDM bounds. We demonstrate this effect in Fig.~\ref{edm_bphyics_unitarity_HB}. It is clear that the B-physics and HiggsBounds impose strong constraints on the parameter space. Furthermore, HiggsBounds imposes the most stringent constraint on the parameter space. 

We show our results in Fig.~\ref{edm_bphyics_unitarity_HB} for two mass ranges for non-standard scalars, $m_{H_i},m_{H^{\pm}}\approx200$ GeV and $m_{H_i},m_{H^{\pm}}\approx 400$ GeV for $i=2,3,4$. We assume $m_{H_1}=125$ GeV, the observed Higgs mass. The scan range for the rest of the parameters is the same as in Table~\ref{scan_relic}. We also maintain the mass difference between the non-standard scalars $\lesssim 50$ GeV in order to satisfy the constraints from S,T,U parameters~\cite{Grimus:2007if,Grimus:2008nb}. 
We can see from Fig.~\ref{edm_bphyics_unitarity_HB} that for the higher masses ($\sim$ 400 GeV) of non-standard scalars, the bounds are relaxed compared to the lower masses of the non-standard scalars ($\sim 200$) GeV.


\section{CP-violating non-standard scalars in the Higgs sector at the future colliders}
\label{collider}

We have identified regions of parameter space in the complex 2HDMS that satisfy the EDM bounds, all experimental and theoretical constraints, and at the same time account for the observed relic density. Now, we ask the question, can we get a hint of CP-violation at the collider? We focus here on the future high-energy $e^+e^-$ collider. 

In this context, future linear colliders~\cite{ECFADESYLCPhysicsWorkingGroup:2001igx,LinearColliderACFAWorkingGroup:2001awr,LinearColliderAmericanWorkingGroup:2001tzv,Abe:2001rdr,Abe:2001grn,LinearColliderAmericanWorkingGroup:2001rzk} are promising next-generation high-energy machines that can directly probe heavy non-standard scalar states.  The linear colliders offer several advantages, e.g. tunable energies and polarized beams~\cite{Moortgat-Pick:2005jsx} and complement current collider projects in exploring the SM and beyond (BSM) \cite{LinearColliderVision:2025hlt,Moortgat-Pick:2015lbx,LHCLCStudyGroup:2004iyd}

We reiterate here that at this point in our study, CP-violation is exclusively in the BSM scalar sector and the 125 GeV Higgs is CP-even and SM-like. In such a scenario, the CP-violation will be manifest in the scalar-trilinear couplings of the non-standard scalars. We investigate this in the following. In the complex 2HDMS model, $H_1$ is the 125 GeV scalar. The non-standard scalar mass-eigenstates are $H_2, H_3, H_4$. In the absence of any mixing, $H_3$ would be a pure pseudo-scalar $a_2$, $H_2$ is a non-standard doublet-like scalar $h_2$, and $H_4$ is a non-standard singlet-like scalar $h_S$. At an $e^+e^-$ collider, the processes we will look into are the following.
\begin{eqnarray}
\label{h2h3}
e^+e^- \rightarrow H_2 H_2 H_3,~~~e^+e^- \rightarrow H_2 H_3 H_3
\end{eqnarray}
and 
\begin{eqnarray}
\label{h4h3}
e^+e^- \rightarrow H_4 H_4 H_3,~~~e^+e^- \rightarrow H_4 H_3 H_3. 
\end{eqnarray}

\begin{figure}[!hptb]
	\centering
        \includegraphics[width=5.0cm,height=4.0cm]{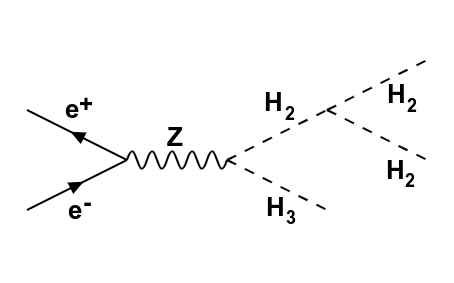}
        \includegraphics[width=5.0cm,height=4.0cm]{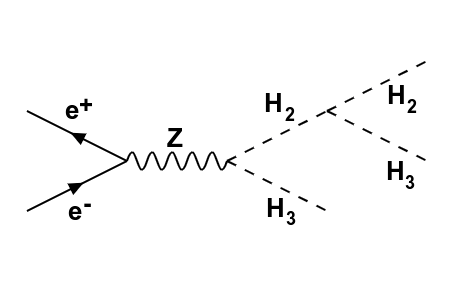} \\
        \includegraphics[width=5.0cm,height=4.0cm]{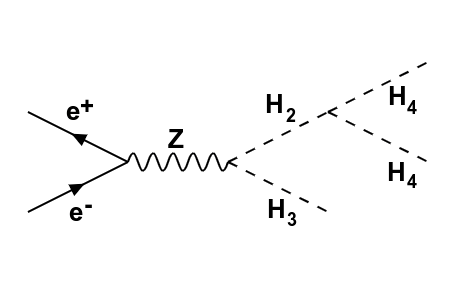}
        \includegraphics[width=5.0cm,height=4.0cm]{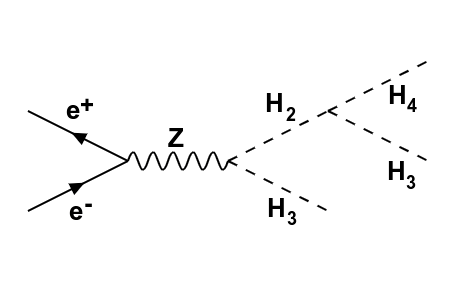}
        \caption{Feynman diagrams for the CP-conserving (left) and CP-violating processes (right).}
        \label{feynman_diagram}
       \end{figure}

\noindent
The Feynman diagram that will have a major contribution to $e^+e^- \rightarrow H_2 H_2 H_3$ is shown in Fig.~\ref{feynman_diagram} (top left).
The diagram with the major contribution to the process $e^+e^- \rightarrow H_2 H_3 H_3$ is shown in Fig.~\ref{feynman_diagram} (top right). On the other hand, 
$e^+e^- \rightarrow H_4 H_4 H_3$ receives dominant contribution from the process shown in Fig.~\ref{feynman_diagram} (bottom left) and $e^+e^- \rightarrow H_4 H_3 H_3$ is dominated by Fig.~\ref{feynman_diagram} (bottom right).

We should note that the coupling of $ZH_3H_2$ is the largest in the alignment limit among the $Z$-couplings to the scalar pairs, and the $Z$-couplings to all other scalar pair combinations are negligible. Therefore, the processes with $H_2 H_3$ intermediate pair play a dominant role, and those are the only combinations shown in Fig.~\ref{feynman_diagram}. Interestingly, it is therefore evident that the imprint of CP-violation will be encoded in the trilinear scalar couplings.

The CP-violating trilinear Higgs couplings in the Higgs basis are as follows:

\begin{eqnarray}
 \label{trilinear_higgsbasis}
    a_2 h_2 h_2 \sim \text{Im}[\lambda_7] v, \\
    a_2 a_2 a_2 \sim \text{Im}[\lambda_7] v, \\
    a_2 h_S h_S \sim \text{Im}[\lambda_6' + 2 \lambda_7'] v.
\end{eqnarray}

\noindent
It is clear from Eq.~\ref{trilinear_higgsbasis} that the couplings $a_2 h_2 h_2, a_2 a_2 a_2$ involve imaginary part of $\lambda_7$ and consequently a non-zero $\theta_7$, whereas $a_2 h_S h_S$ involves imaginary part of $\lambda_6' + 2 \lambda_7'$, thereby a nonzero $\theta_{CP}$. 
Evidently, in the absence of CP-violation, these vertices do not exist.

The CP-conserving trilinear Higgs couplings in the Higgs basis are as follows:

\begin{eqnarray}
 \label{trilinear_higgsbasis}
    h_2 h_2 h_2 \sim \text{Re}[\lambda_7] v, \\
    h_2 a_2 a_2 \sim \text{Re}[\lambda_7] v, \\
    h_2 h_S h_S \sim \text{Re}[\lambda_6' + 2 \lambda_7'] v.
\end{eqnarray}

\noindent
The couplings $h_2 h_2 h_2, h_2h_Sh_S$ (three scalars),  and $h_2 a_2 a_2$ (scalar-pseudoscalar-pseudocalar) couplings are proportional to the real part of the couplings $\lambda_7$ and $\lambda_6' + 2 \lambda_7'$ respectively, and therefore are non-zero in the CP-conserving case. 

In the absence of CP-violation, $H_3\rightarrow a_2$, $H_2\rightarrow h_2$ and $H_4\rightarrow h_S$. Therefore, in the CP-conserving limit, only one of the two processes $e^+ e^- \rightarrow H_2 H_2 H_3$ and $e^+ e^- \rightarrow H_2 H_3 H_3$ will take place. But in the CP-violating case, both processes will be present. The same argument will follow in case of the $e^+ e^- \rightarrow H_4 H_4 H_3$ and $e^+ e^- \rightarrow H_4 H_3 H_3$ final states. We reiterate that $e^+ e^- \rightarrow H_2 H_3 H_3$ and $e^+ e^- \rightarrow H_4 H_3 H_3$ processes will be forbidden from CP-conservation, but those processes will be present along with $e^+ e^- \rightarrow H_2 H_2 H_3$ and $e^+ e^- \rightarrow H_4 H_4 H_3$, in case of CP-violation.


\noindent       
Therefore, as we discussed, the CP-violation at the collider can be probed through CP-violating trilinear couplings via simultaneous observation of the following processes: 
      \begin{enumerate}
      \item $e^+ e^- \rightarrow H_2 H_2 H_3,$ 
      \item $e^+ e^- \rightarrow H_2 H_3 H_3,$     \item $e^+ e^- \rightarrow H_4 H_4 H_3,$ and 
      \item $e^+ e^- \rightarrow H_4 H_3 H_3.$ 
      \end{enumerate}   

\noindent
  Furthermore, CP-violating Process (2) $e^+ e^- \rightarrow H_2 H_3 H_3$ is particularly sensitive to $\theta_7$, Process (4) $e^+ e^- \rightarrow H_4 H_3 H_3$  sensitive to $\theta_{CP}$. However, the dependence on the phases is not straightforward, since $H_2, H_3, H_4$ are the mass eigenstates and the CP-even and CP-odd parts in them are governed by the mixing angles $\alpha_{1,2,3}$.


\begin{flushleft}
\begin{table}
{\scriptsize{\begin{tabular}{|c|c|c|c|c|c|c|c|c|c|c|c|}
\hline
BP's  & $m_{H_2}$ & $m_{H_3}$ & $m_{H_4}$ & $M$ & $m_H^{\pm}$ & $v_S$ & Re$[\lambda_6'+2\lambda_7']$ & Im$[\lambda_6'+2\lambda_7']$ & Re$[\lambda_7]$ & Im$[\lambda_7]$ & [$\alpha_1,\alpha_2,\alpha_3$] \\
\hline
BP1 &  200 & 230 & 250 & 240 & 230 & 100 & -0.76 & 0 & 0.12 & 0 & [0,-0.9,0] \\
\hline
BP2 &  200 & 230 & 250 & 240 & 230 & 100 & 0.45 & 0.5 & 0.25 & 0.24 & [1.2,0.4,-0.4] \\
\hline
BP3 &  400 & 427 & 450 & 400 & 427 & 200 & -0.7 & 0 & 1.3 & 0 & [0,0.9,0] \\
\hline
BP4 &  400 & 427 & 450 & 400 & 427 & 200 &0.44 & 0.5 & 1.14 & 1.13 & [0.7,0.3,-0.6] \\
\hline
\end{tabular}}}
\caption{The chosen benchmarks that satisfy all theoretical and experimental constraints, including the EDM bounds, account for dark matter. BP1 and BP3 correspond to CP-conserving scenarios, while BP2 and BP4 are CP-violating. In all cases $M_{H_1}$ is 125 GeV, {\it i.e.} the observed Higgs mass.}
\label{benchmarks}
\end{table}
\end{flushleft}

\noindent
 We chose a benchmark with masses of the non-standard heavy scalars around 200 GeV (BP1, BP2) and 400 GeV (BP3, BP4), shown in Table~\ref{benchmarks}. The benchmarks satisfy all constraints, including the EDM bound and account for the observed relic density. In two of the chosen BP's, namely BP1 and BP3, we put all the CP-violating phases to zero, i.e $\theta_{CP}=0, \theta_7=0$ (which implies Im$[\lambda_6'+2\lambda_7']$ = Im$[\lambda_7] = 0$) and $\theta_{u,d,e}$ = 0.  We calculate the cross-section of the aforementioned four processes as a function of $\sqrt{s}$ in this case using \texttt{MG5$\_$aMC$\_$v3.5.5}~\cite{Alwall:2014hca,Alwall:2011uj}. 
 
 \begin{figure}[!hptb]
        \includegraphics[width=7.0cm,height=6.0cm]{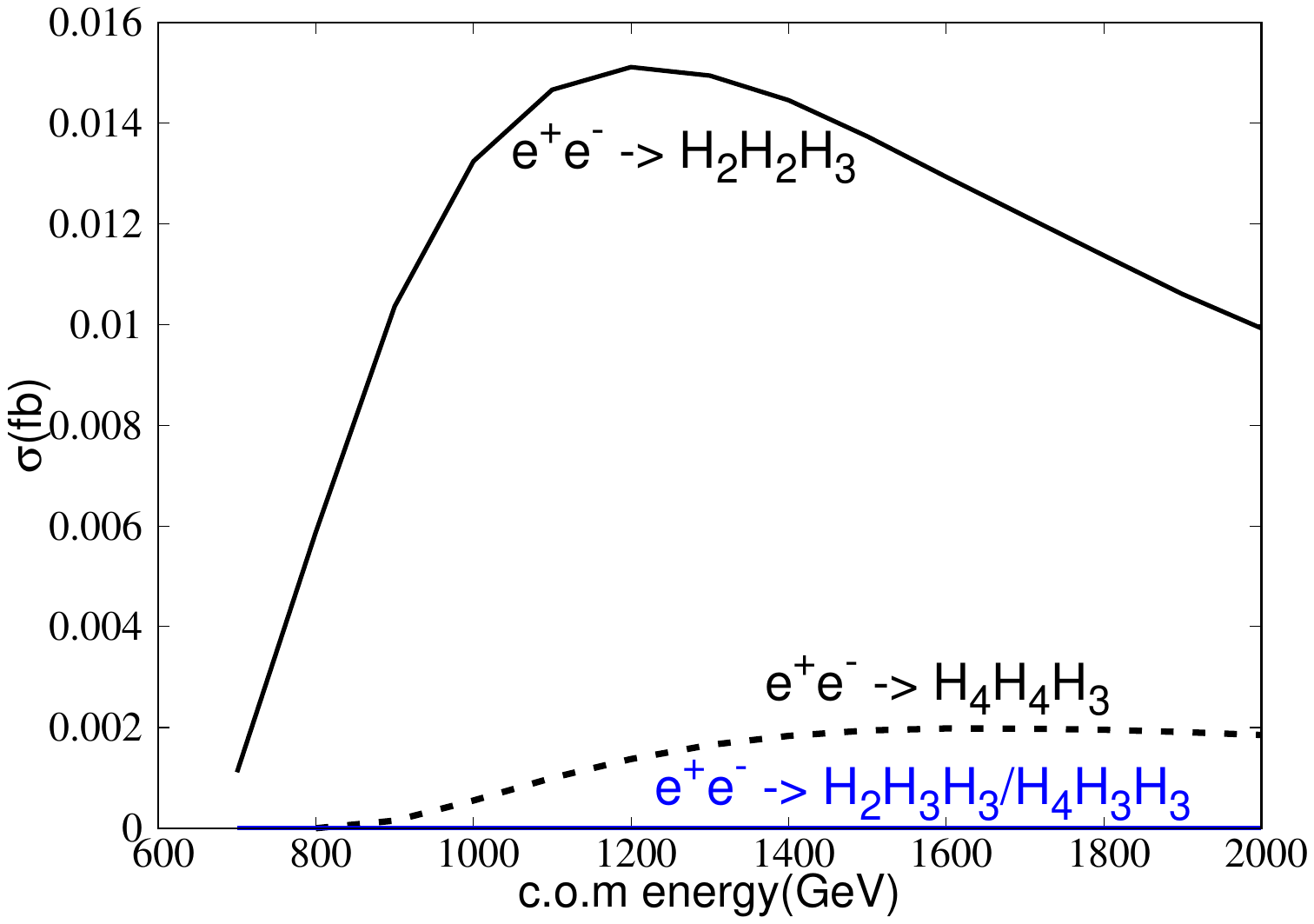}
        \hspace{1.0cm}
     \includegraphics[width=7.0cm,height=6.0cm]{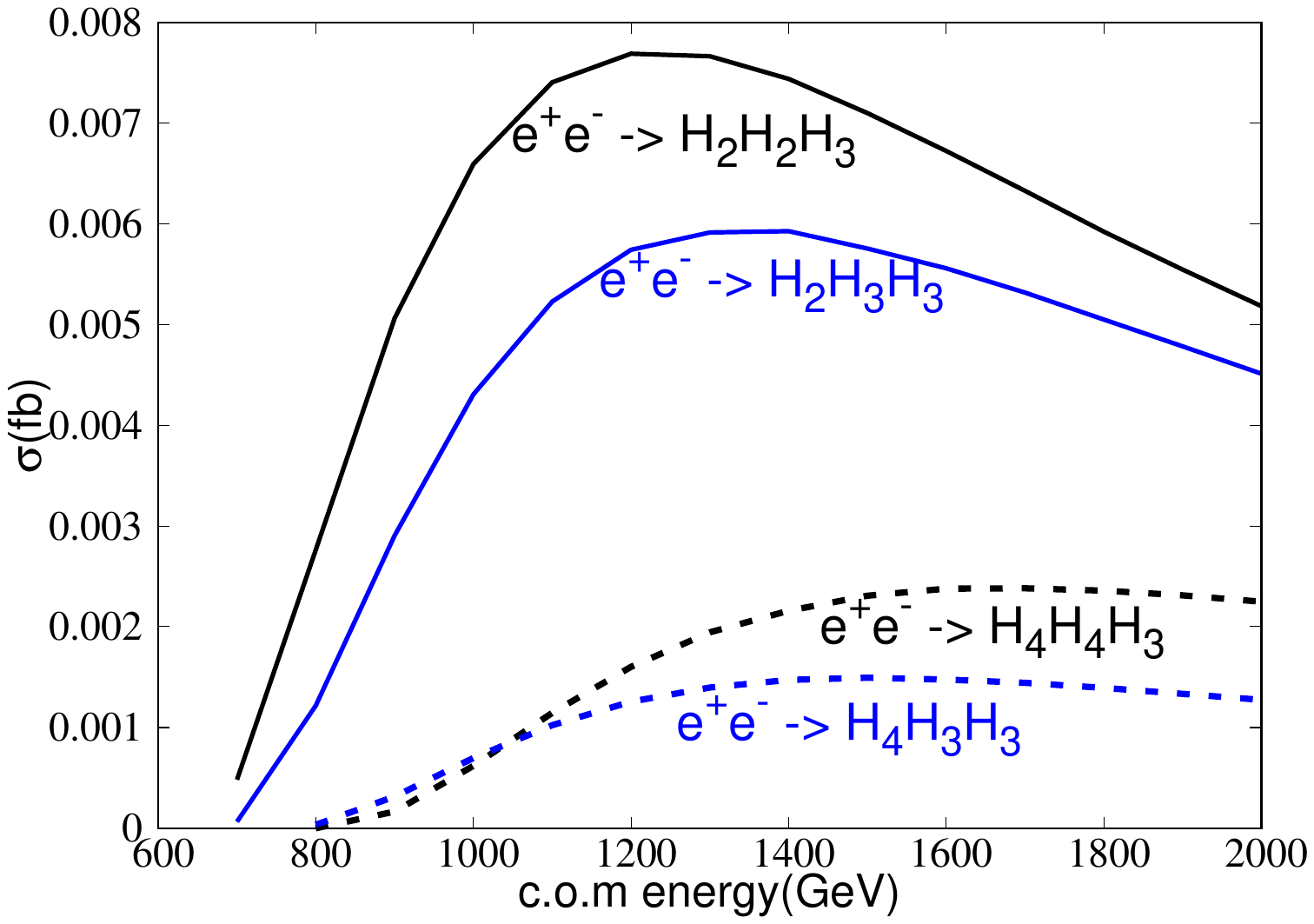} \\
        \caption{Cross-sections for the CP-conserving and CP-violating processes for BP1 (left) and BP2 (right) as a function of $\sqrt{s}$.}
        \label{crosssections_200}
\end{figure}

\begin{figure}[!hptb]
    \includegraphics[width=7.0cm,height=6.0cm]{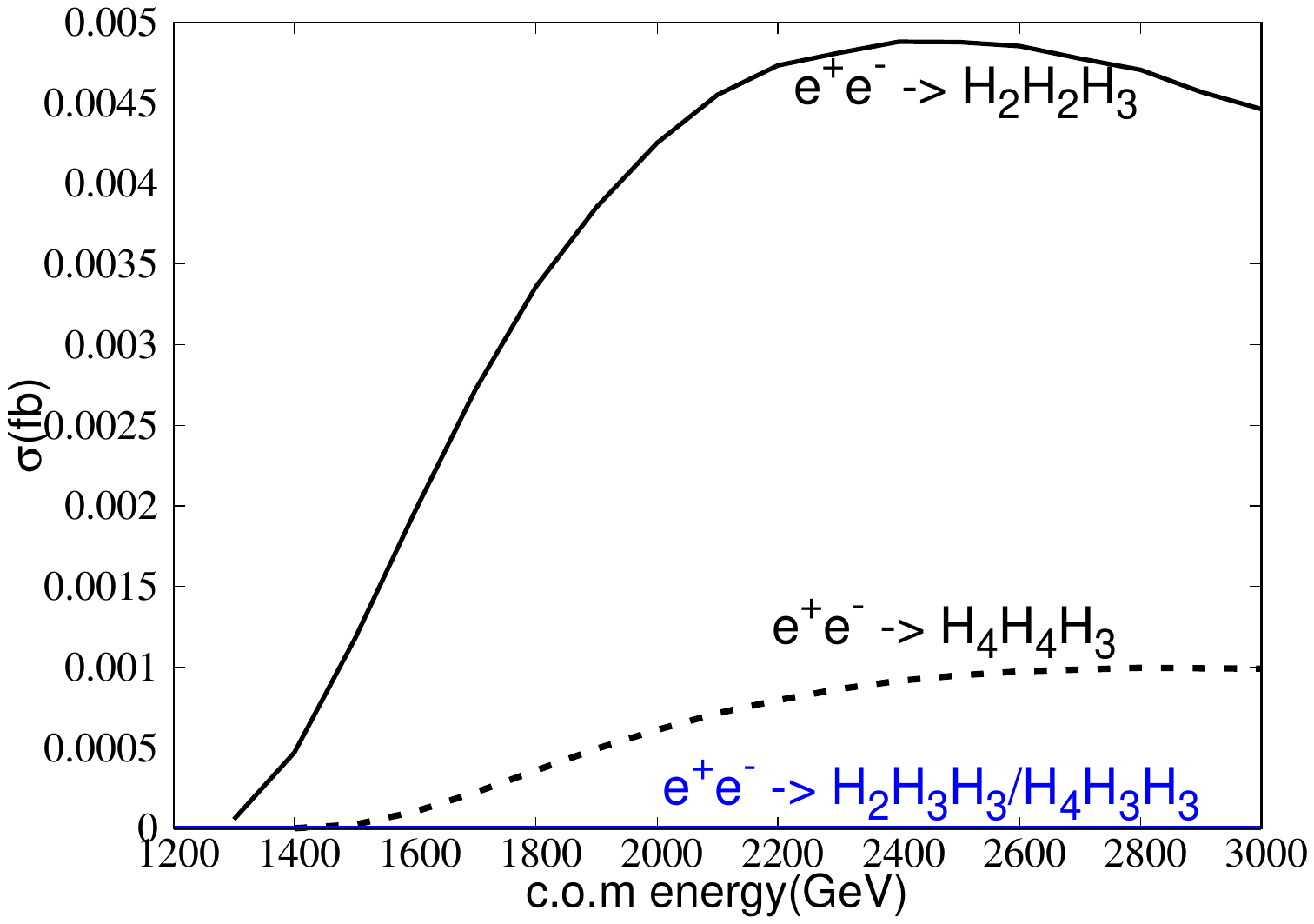}
    \hspace{1.0cm}
    \includegraphics[width=7.0cm,height=6.0cm]{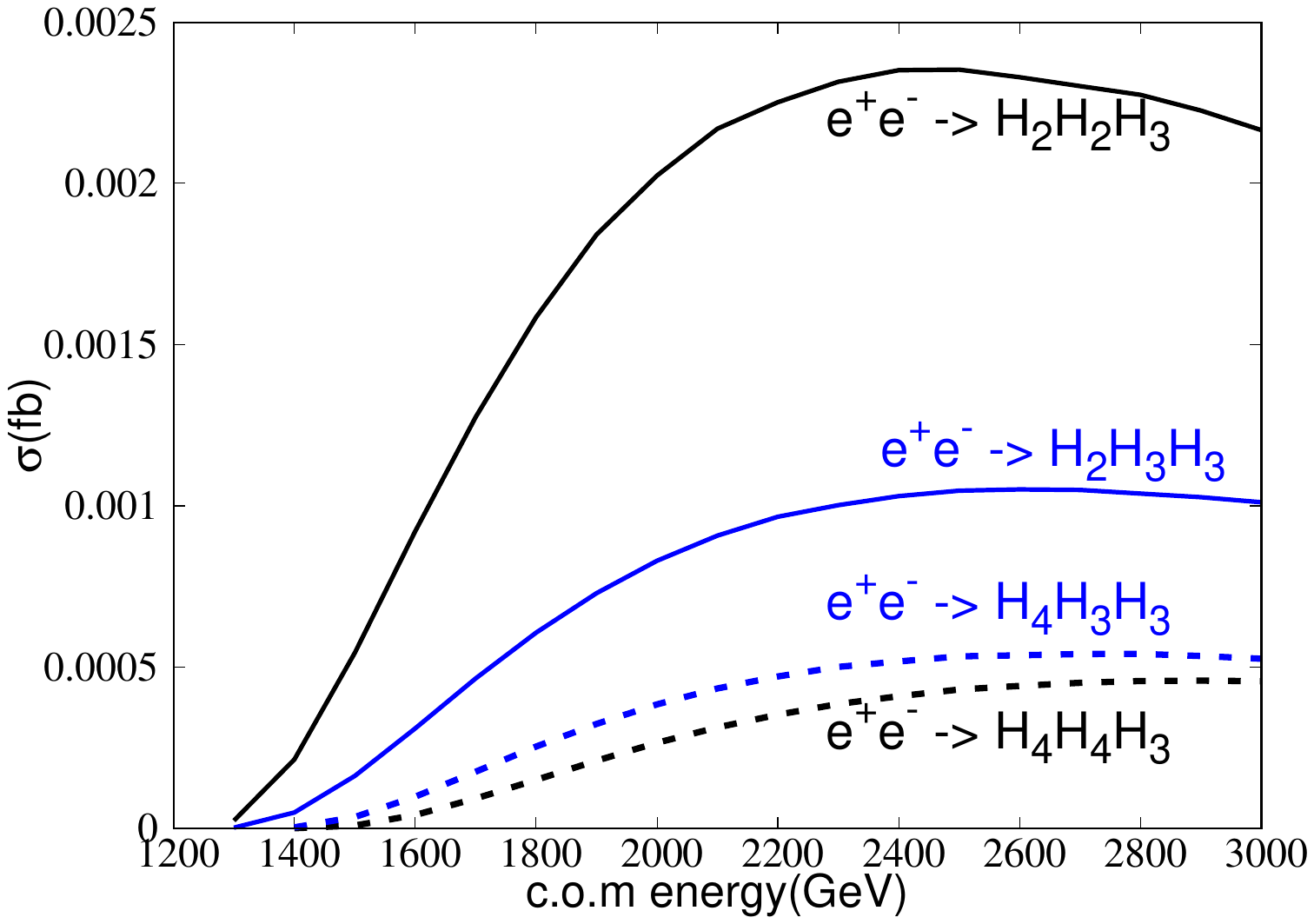} \\
        \caption{Cross-sections for the CP-conserving and CP-violating processes for BP3 (left) and BP4 (right) as a function of $\sqrt{s}$.}
        \label{crosssections_400}
\end{figure}

 One can see in Fig.~\ref{crosssections_200} and ~\ref{crosssections_400} (left panel), only the CP-conserving processes, namely $e^+ e^- \rightarrow H_2 H_2 H_3$ (black solid curve) and $e^+ e^- \rightarrow H_4 H_4 H_3$ (black dashed curve) will have non-zero cross-section, while the cross-sections for the CP-violating processes, namely $e^+ e^- \rightarrow H_2 H_3 H_3$ and $e^+ e^- \rightarrow H_4 H_3 H_3$ (blue curves) are zero. Next, we choose two other benchmarks, BP2 and BP4, where the CP-violation is non-zero, which can be seen from non-zero values of Im$[\lambda_6'+2\lambda_7']$ and Im$[\lambda_7]$ in Table~\ref{benchmarks}. The phases in the Yukawa sector, $\theta_{u,d,e}$, are also non-zero, and they are chosen from the scan such that the EDM bounds are satisfied for both benchmarks. However, the phases in the Yukawa sector are not particularly relevant for the process under current consideration. In the right panels of Figs.~\ref{crosssections_200} and ~\ref{crosssections_400}, we can see that, in the presence of non-zero CP-phases, the CP-violating processes  $e^+ e^- \rightarrow H_2 H_3 H_3$ (solid blue) and $e^+ e^- \rightarrow H_4 H_3 H_3$ (dashed blue) yield non-zero cross-section. We also note that a further enhancement of the cross-sections is possible using polarized beams for both electron and positron. We present in Table~\ref{polarized} the comparison between unpolarized and polarized cross-sections of the process $e^+e^- \rightarrow H_2 H_3 H_3$ in the case of BP2 at $\sqrt{s}$=1.5 TeV and for the process $e^+e^- \rightarrow H_4 H_3 H_3$ in the case of BP4 at $\sqrt{s}$=3 TeV. We chose the proposed CLIC design energies 1.5 TeV and 3 TeV for our purpose~\cite{CLICdp:2018cto}. Furthermore, in both cases, we chose an optimistic polarization choice for electron and positron beams ($P^{e^-} = \pm$ 90\% and $P^{e^+} = \pm$ 60\%). It can be seen from Table~\ref{polarized}, that a factor 1.34 enhancement is possible when $P^{e^-} = $ 90\% and $P^{e^+} = $ -60\%, and a factor of 1.7 can be achieved with $P^{e^-} = $ -90\% and $P^{e^+} = $60\% for both final states. We mention that our choice of polarization of the $e^-$ and especially $e^+$ is optimistic and larger compared to the design polarization of CLIC~\cite{CLICdp:2018cto}.

A few comments are in order on the typical cross-sections achievable for the aforementioned processes. We have discussed earlier the matrix elements as a function of the coupling parameters. One can rewrite those expressions in terms of masses and rotation matrix elements ($R_{ij}$'s) as follows.

\begin{eqnarray}
        m_{34} &=& \text{Im}[\lambda_6'+2 \lambda_7'],\\
         m_{34} &=& R_{32}R_{42}m_{H_2}^2 + R_{33}R_{43}m_{H_3}^2 + R_{34} R_{44} m_{H_4}^2. 
 \label{sumrule}    
\end{eqnarray} 
From the sum rule above(Eq.~\ref{sumrule}), one can see that the trilinear CP-violating coupling, which is proportional to $\text{Im}[\lambda_6'+2 \lambda_7']$, is bounded from above because of the unitarity of the rotation matrix $R_{ij}$. We have used the maximum value that can be achieved for the CP-violating trilinear coupling allowed by the sum-rule, while simultaneously satisfying all constraints as well as the observed relic density in both benchmarks BP2 and BP4. On the other hand, on the coupling $\text{Im}[\lambda_7]$, no similar bound exists because $\lambda_7$ coupling does not take part in the mass mixing and therefore no similar sum rule (as in Eq.~\ref{sumrule}) applies. We abide by the perturbativity condition on the $\lambda_7$ coupling.

\begin{table}
        \centering
      \label{multiprogram}
        \begin{tabular}{|c|c|c|c|c|c|}
        \hline
       &  &   & \multicolumn{3}{|c|}{Cross-sections (fb)}\\
        \cline{3-6}
       Benchmark & Process & $\sqrt{s}$ &  Unpolarized & $P^{e^-}$ = 90\% & $P^{e^-}$ = -90\%  \\
        &  &  & &$P^{e^+}$ = -60\% & $P^{e^+}$ = 60\% \\
            \cline{1-6} 
             BP2 & $e^+e^-\rightarrow H_2 H_3 H_3$ & 1.5 TeV&0.0058 & 0.0078 & 0.01   \\
         \hline
         BP4 & $e^+e^-\rightarrow H_4 H_3 H_3$ & 3 TeV &0.00053 & 0.00071 & 0.00092 \\
            \hline
        \end{tabular}
         \caption{Enhancement of cross-sections with polarized beams.}
         \label{polarized}
    \end{table}

\section{Deviation from the exact alignment and CP-phase of the 125 GeV Higgs boson}
\label{collider125}

One of the primary goals of the present and future colliders is the study of the Higgs boson properties.
So far, we have assumed in our analysis the exact alignment limit, which means the observed 125 GeV Higgs boson~\cite{ATLAS:2012yve,CMS:2012qbp} is exactly SM-like. In such scenarios, the CP-violation is assumed to be exclusively in the BSM sector. This is a viable possibility, keeping in mind the experimental data favor a CP-even 125 GeV Higgs boson~\cite{ATLAS:2020ior,ATLAS:2022akr,ATLAS:2022tan}. Having said that, we should not ignore the fact that there is still room for exploration in the CP properties of the 125 GeV scalar~\cite{Yang:2025gyx}. In the future runs of the LHC as well as future lepton colliders such as FCC-ee~\cite{Abada:2019zxq,Agapov:2022bhm,Ge:2020mcl}, CEPC~\cite{CEPCStudyGroup:2018ghi,CEPCStudyGroup:2018rmc,Sha:2022bkt} (known as Higgs factories), it would be possible to probe the CP nature of the observed Higgs boson even more precisely. 
Linear colliders, such as the International Linear Collider (ILC) \cite{Bambade:2019fyw}, or the Compact Linear Collider (CLIC) \cite{CLICdp:2018cto}, that are designed to operate mainly at centre-of-mass energies between 250 GeV and 3 TeV. This energy range is suitable for producing Higgs bosons and studying their properties, including the trilinear Higgs coupling, which is essential for probing the structure of the Higgs potential \cite{deBlas:2019rxi, Bechtle:2024acc, LinearColliderVision:2025hlt}.
Therefore, we will now deviate from our approach of exact alignment condition and allow larger CP-mixing in the scalar sector, which will lead to CP-violating interactions of the 125 GeV Higgs. 

The mass matrix in the Higgs basis, in this case, will be the following. 

\begin{equation*}
{\cal{M}}_{ij}^2=
\left(
\begin{array}{cccc|c} 
  M_{11} & M_{12} & M_{13} & M_{14} & 0 \\  
  M_{12} & M_{22} & 0 & M_{24} & 0 \\
  M_{13} & 0 & M_{33} & M_{34} & 0 \\
  M_{14} & M_{24} & M_{34} & M_{44} & 0 \\
  \hline  
  0 & 0 & 0 & 0 & M_{55} \\
\end{array} 
\right),
\end{equation*}
where
\begin{eqnarray}
M_{12} &=& \text{Re}[\lambda_6] v^2, \nonumber \\
M_{13} &=& -\text{Im}[\lambda_6] v^2,\nonumber \\
M_{14} &=& v v_S (\lambda_1' + 2 \text{Re}  [\lambda_4']).
\label{matrixelements_noalignment}
\end{eqnarray}
The rest of the matrix elements are the same as in Eqs.~\ref{matrixelements_alignment}.

It is clear from Eq.~\ref{matrixelements_noalignment} that the phase of $\lambda_{6}$ will introduce CP-mixing between the pseudoscalar and the scalar state, and the 125 GeV scalar will be a CP-mixed mass eigenstate. 

\subsection{Impact of CP-violation in the Yukawa coupling of the 125 GeV Higgs}

 The aforementioned CP-mixing in the 125 GeV scalar ($h$) through mass mixing, will strongly manifest in its CP-violating Yukawa couplings. If we go back to the Yukawa interactions of the scalars in this model in Eq~\ref{yukawa}, we can calculate the scale factor $\kappa_f^1$ of the Yukawa coupling of the 125 GeV scalar. This scale factor becomes 1 in the exact alignment limit as discussed before. However, in the most general case, the scale factor becomes 

\begin{eqnarray}
			\mathcal{L}_\text{yukawa}&=&-\sum_{f=u,d,e}\cbra{\bar{f}_L M_f f_R +\sum_{j=1}^3\bar{f}_L \rbra{\frac{M_f}{v} \kappa_f^j}f_R H^0_j+h.c.},\\
\kappa_f^1&=&\mathcal{R}_{11}+\sbra{\mathcal{R}_{21} +i(-2I_f)\mathcal{R}_{31}} |\zeta_f|e^{i(-2I_f)\theta_f}.
\end{eqnarray}

\noindent
The Yukawa interaction of the 125 GeV scalar can be written as follows:
\begin{equation}
    \mathcal{L}_{hff}= \frac{M_f}{v}|\kappa_f|\bar{f}_L(\cos\xi_f + i \gamma_5 \sin\xi_f)f_R h.
\end{equation}

\noindent
The angles $\xi_f$ would be the measure of CP-violation in the 125 GeV Higgs, pertaining to its coupling to various fermions. LHC experiments have made measurements of the fermionic coupling of the Higgs boson and have also looked at CP-sensitive observables to constrain the phase $\sin\xi_f$. The most recent and strongest limits come from ~\cite{CMS:2022uox,ATLAS:2022akr,ATLAS:2023cbt}. Furthermore, there are future experiments, such as the proposed lepton colliders, which can constrain these angles further. Also, if such a CP-violating phase is present in nature, future lepton colliders will be able to probe it with remarkable precision~\cite{Sha:2022bkt}. However, one should also remember that such phases will have an impact on the EDM calculation too and receive strong constraints from the EDM measurements. 
Our goal is to see if the phases that are likely to be probed in future experiments, are allowed by the EDM experiments. 

\begin{figure}[!hptb]
	\centering
    \includegraphics[width=5.5cm,height=4.5cm]{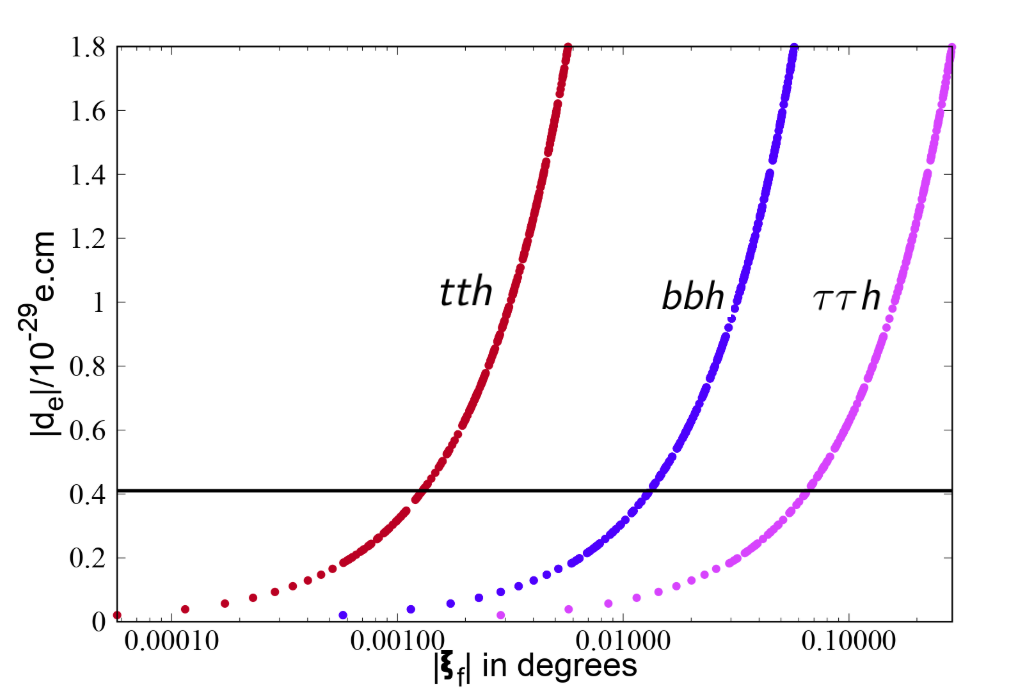}
    \caption{The electron EDM as a function of the effective CP-violating phases in the 125 GeV scalar in the $tth$, $bbh$, and $\tau\tau h$ coupling. $|\kappa_f|\approx 1$ for all three cases.}
    \label{angle_edm}
\end{figure}

\noindent
We found that the answer to this question is model-dependent. We first consider one of the extreme scenarios. We assume the only CP-violating phase is in the CP-mixing in the 125 GeV scalar. In other words, only $\text{Im}[\lambda_6] \neq 0$ and all other parameters are real. In such a scenario, due to the presence of only one phase, there will be no possibility of cancellation between various loop contributions and the CP-violating angles $\xi_f$ will receive strong constraints from the EDM bounds. We show this in Fig.~\ref{angle_edm}. 

One can clearly see that the phase $\xi_f$ gets strongly constrained by the EDM data. The limits are strongest for the $t\bar t h$ coupling, since the top-loop contribution is the largest in the two-loop Bar-Zee diagrams. It is also clear by the comparison with future projections of the lepton colliders for CP-violating phases, that if the CP-phase is present only in the 125 GeV Higgs boson, then it has to be extremely small to obey the EDM bounds. Therefore, it will not be possible to probe such a phase at future lepton colliders. 
However, with other nonzero phases present in the model, relative cancellation between diagrams can lead to less restrictive bounds on $\xi_f$ from EDM experiments. Therefore, we reiterate that future probes of CP-violation in future collider runs become model-dependent. 

\begin{figure}[!hptb]
	\centering
    \includegraphics[width=6.5cm,height=5.5cm]{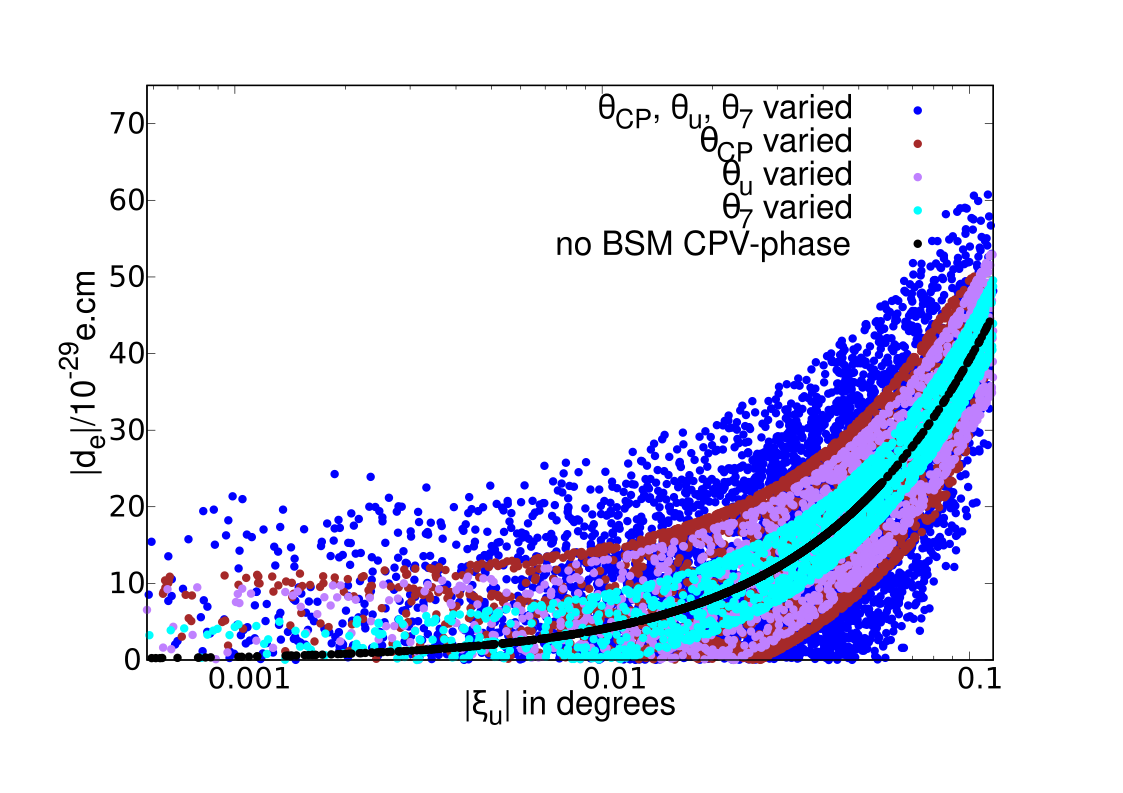}
    \includegraphics[width=6.5cm,height=5.5cm]{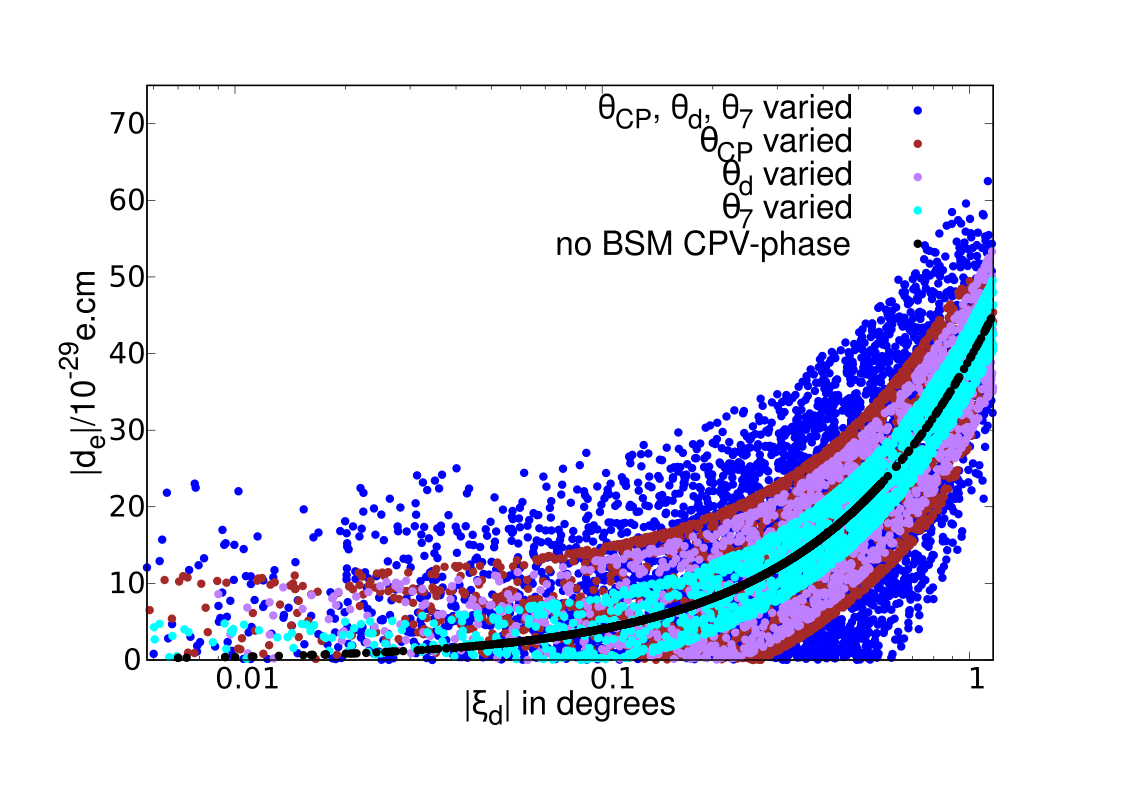}  
    \includegraphics[width=6.5cm,height=5.5cm]{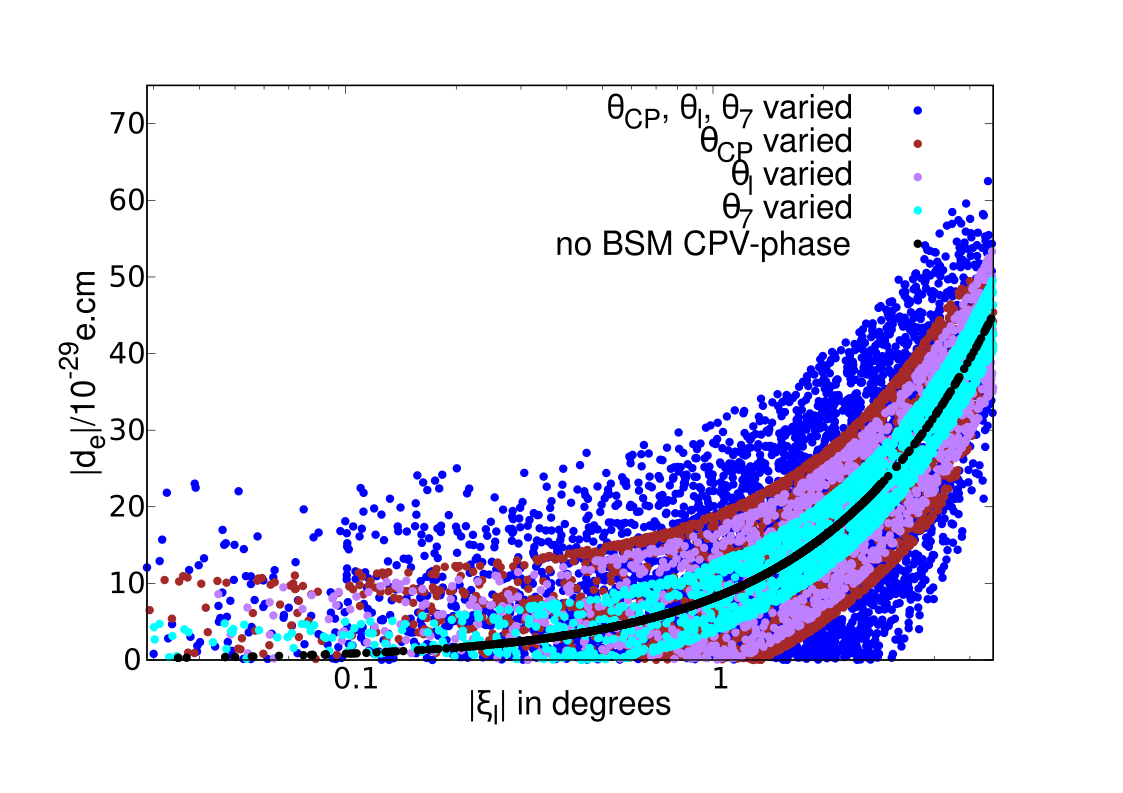}
    \caption{The electron EDM as a function of the effective CP-violating phases in the 125 GeV scalar in the $tth$, $bbh$ and $\tau\tau h$ coupling.}
    \label{angle_edm_varied}
\end{figure}

We show next the impact of non-zero phases in the BSM sector in our model on the limits on the effective CP-violating Yukawa phases of the 125 GeV Higgs. 
We study the impact of the sources of CP-violation in the BSM sector, on the limit on $\xi_u, \xi_d$ and $\xi_l$ obtained earlier.

We first vary the three phases $\theta_{CP}, \theta_u$ and $\theta_7$ one at a time. Next, we vary three phases together. We show the results in Fig.~\ref{angle_edm_varied}. Fig.~\ref{angle_edm_varied} (top left) shows the limit on the CP-violating $t\bar t h$ coupling. Fig.~\ref{angle_edm_varied} (top right) shows the limit on CP-violating $b\bar b h$ coupling, and the limits on CP-violating $\tau\tau h$ coupling are shown in Fig.~\ref{angle_edm_varied} (centre). The black curves in the three aforementioned plots are the same as the red, blue and purple curves in Fig.~\ref{angle_edm}. This shows that when all the phases in the BSM sector of 2HDMS are kept zero {\it i.e.} $\theta_{CP}=\theta_{7}=\theta_u\approx 0$, the limits on the CP-violating couplings of the 125 GeV scalars are strongly constrained by the EDM bounds. However, when we start varying the phases in the BSM sector, the scenario changes. In Fig.~\ref{angle_edm_varied}, the cyan region shows the variation of electron EDM $|d_e|$ as a function of $\xi_f$ when only $\theta_7\neq 0$ and $\theta_{CP}=\theta_u=0$. We can see the limit on $\xi_f$ is relaxed compared to the case where all the BSM phases are zero (black curve). The purple region shows the variation of $|d_e|$ as a function of $\xi_f$ when $\theta_u$ is varied, and $\theta_{CP}=\theta_7=0$. The brown region is obtained when $\theta_{CP}$ is varied, and the other two phases $\theta_7$ and $\theta_u$ are kept zero. In all three cases, we see that the bounds on $\xi_f$ are relaxed compared to the case where all three phases were zero. This happens due to the cancellation between various diagrams involving the 125 GeV scalar and diagrams involving the additional scalars. The chances of cancellation are maximal when three phases $\theta_{CP},\theta_7$ and $\theta_u$ are all varied simultaneously. This is shown in the blue regions. We note that the masses of the extra scalars are fixed at 200 GeV to enable effective cancellation. If the scalar masses are high, the cancellation would be less effective, and the limits on $\xi_f$ would be stronger.
We list here the limits on $\xi_u, \xi_d$ and $\xi_l$ in the aforementioned cases in Table~\ref{edmlimits}. One can see that in the presence of all the phases, it may be possible to probe the CP-violating fermion couplings of the 125 GeV Higgs, especially in the $\tau\tau h$ interaction ($\xi_l \sim$ few degrees), at the future lepton colliders~\cite{Giappichini:2026uly}.

\begin{table}[h!]
\centering
\begin{tabular}{|c|c|c|c|c|c|} 
\hline
\multicolumn{3}{|c|}{\textbf{CP-violating phases}} & \multicolumn{3}{c|}{\textbf{EDM Limit on $|\xi_f|$ }}\\
\hline
$\theta_{CP}$ & $\theta_u$ & $\theta_7$&  $\xi_u$ & $\xi_d$ & $\xi_l$ \\
\hline
0& 0 & 0 & 0.001 & 0.01 & 0.1 \\
\hline
0 & 0 & $\neq$ 0 & 0.01 & 0.1 & 0.8\\
\hline
0 & $\neq$ 0 & 0 & 0.02 & 0.2 & 1.2 \\
\hline
$\neq$ 0 & 0 & 0 & 0.03 & 0.3 & 1.3 \\
\hline
$\neq$ 0 & $\neq$ 0 & $\neq$ 0 & 0.08 & 0.7 & 5 \\
\hline
\end{tabular}
\caption{EDM Limits on $|\xi_f|$ in degrees, with various conditions on the CP-violating phases in our model.}
\label{edmlimits}
\end{table}

\noindent
A few comments are in order. In the fermion coupling of the 125 GeV Higgs boson, CP-violation enters via two sources. One is the phase of the parameter $\text{Im}(\lambda_6)$ (see Equation~\ref{matrixelements_noalignment}) and the other is the phases ($\theta_f$) of the Yukawa factors ($\zeta_f$). The effective phase $\xi_f$ discussed is a combined effect of both phases. 

\subsection{Impact of CP-violation on the triple-Higgs coupling of 125 GeV Higgs}

\begin{figure}[!hptb]
	\centering
 \includegraphics[width=7.5cm,height=6.5cm]{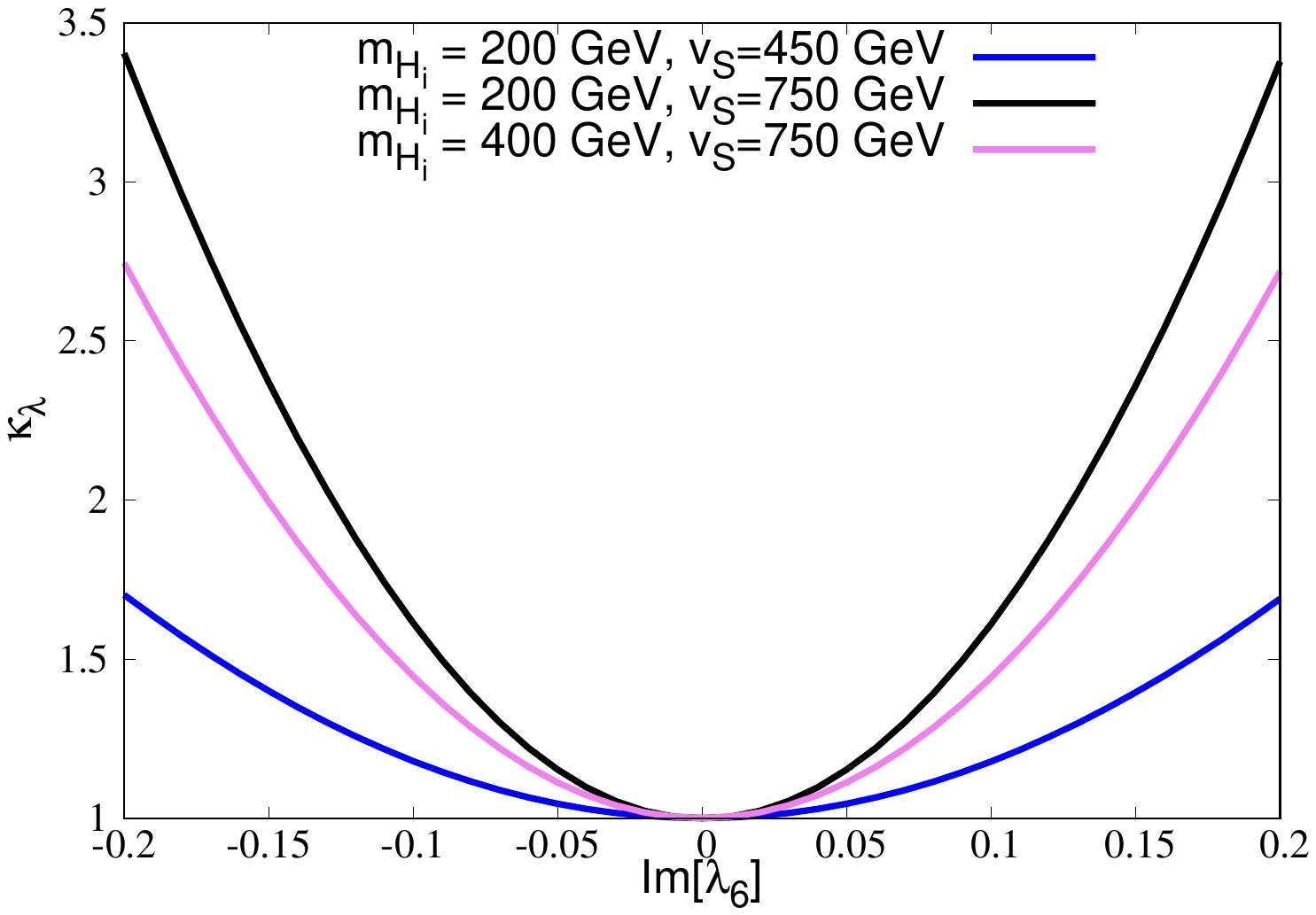}
    \caption{The variation of $\kappa_{\lambda} = \frac{\lambda_{hhh}}{\lambda_{hhh}^{SM}}$ for three benchmark scenarios as a function of $\text{Im}[\lambda_6]$. We chose $\text{Re}[\lambda_6]=0$,$|\lambda_5'|=-1$ and $\lambda_2'=0$ in all cases.}
    \label{trilinear_CPV}
\end{figure}

 After studying the CP-violation in the Yukawa sector, we explore the effect of CP-violation on the trilinear coupling of the 125 GeV Higgs boson.
It is interesting to note that this coupling is CP-sensitive since $\text{Im}[\lambda_6]$ introduces mixing between the CP-odd scalar and the SM Higgs and controls the relative sizes of the CP-even and CP-odd admixture of the 125 GeV Higgs. We confirm that the trilinear Higgs coupling is not sensitive to any other phases in our model, namely, $\theta_{CP}, \theta_f$ or $\theta_7$. We first show the variation of the Higgs trilinear coupling normalised to its Standard Model value $\left(\lambda_{hhh}^{SM}=\frac{3m_h^2}{v}\right)$ as a function of the imaginary part of $\lambda_6$, in Fig.~\ref{trilinear_CPV}. While doing so, we have chosen the mixing angle in such a way that the real part of $\lambda_6$ is 0 (see Eq.~\ref{matrixelements_noalignment} for reference).

\begin{figure}[!hptb]
	\centering
    \includegraphics[width=7.5cm,height=6.5cm]{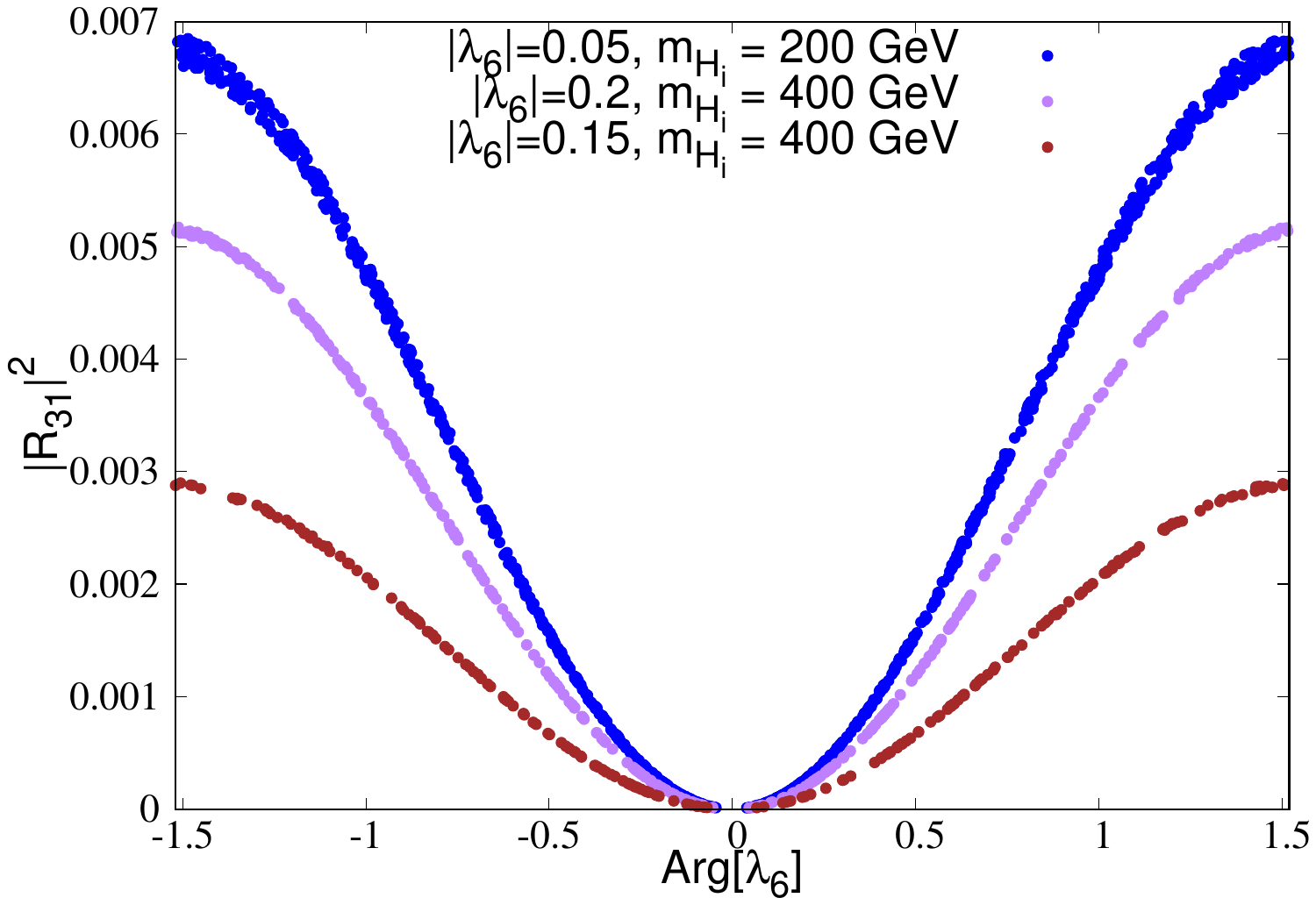}
    \caption{The variation of the CP-odd admixture $|R_{31}|^2$ of the Higgs boson as a function of the phase of $\lambda_6$(namely $\theta_6$). Benchmark values of $|\lambda_6|$ and the non-standard scalar masses are mentioned in the plot.}
    \label{angle_R31}
\end{figure}

\begin{figure}[!hptb]
	\centering
    \includegraphics[width=7.0cm,height=6.0cm]{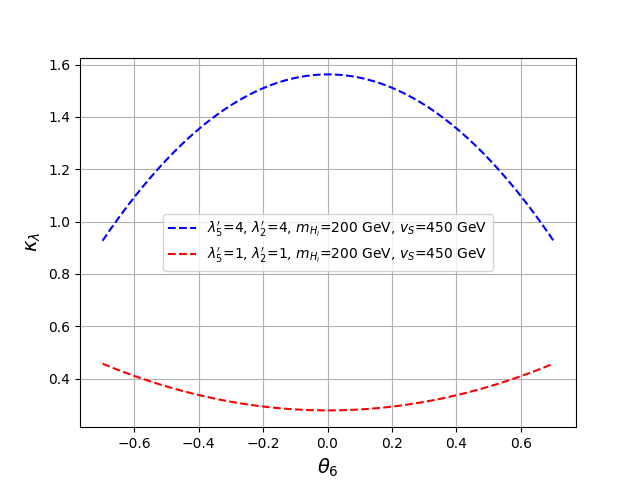}
 \includegraphics[width=7.0cm,height=6.0cm]{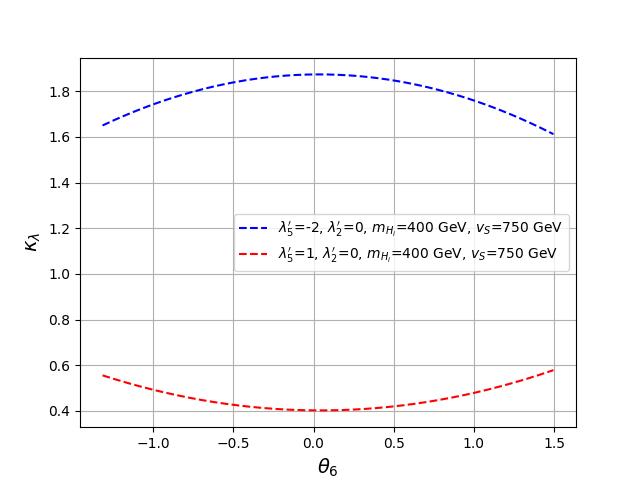}
    \caption{Variation of $\kappa_\lambda$ as a function of the CP-violating phase of $\lambda_6$ ($\theta_6$), while $|\lambda_6| $ is fixed at 0.05 (left) and 0.2 (right). The chosen values of the other relevant parameters are also shown.}
    \label{trilinear}
\end{figure}

\noindent
We have chosen the range of $\text{Im}[\lambda_6]$ such that the corresponding benchmarks satisfy the EDM bounds. All the other aforementioned theoretical and experimental constraints are also taken into account. One can see the SM limit is retrieved when $\text{Im}[\lambda_6]=0$. Also, even at the tree-level, obtaining a large deviation from the SM prediction of triple-Higgs coupling is possible in the presence of CP-violation.

Next, we will try to understand the impact of the phase of $\lambda_6$ on the triple-Higgs coupling. For that purpose, we will keep the $|\lambda_6|$ fixed. We show in Fig.~\ref{angle_R31}, the variation of the CP-odd admixture $|R_{31}|^2$ of the Higgs boson as a function of the phase of $\lambda_6$(namely $\theta_6$). We have chosen three different values of $|\lambda_6|$ as well as two benchmark non-standard scalar masses for our analysis. We can see that the CP-odd admixture increases with decreasing mass of the non-standard scalar masses. Also, a larger $|\lambda_6|$ will introduce a bigger CP-odd admixture.

Now, we study the variation of the Higgs trilinear coupling parametrised by $\kappa_\lambda$, as a function of $\text{Arg}[\lambda_6]$ or $\theta_6$, when $|\lambda_6|$
is fixed. We can show this in Fig.~\ref{trilinear}. On the left, we chose $|\lambda_6|=0.05$, and for the plot on the right, $|\lambda_6|=0.2$. We can see from the plots that the dependence on $\theta_6$ is non-trivial and depends largely on the other parameters. We have already seen in Fig.~\ref{trilinear_CPV}, the impact of the non-standard scalar masses $m_{H_i}$ and the singlet vev $v_S$ on the self-coupling. We can see that effect here, too. Furthermore, the couplings $\lambda_2',\lambda_5'$ also play a deciding factor. The nature of the variation of $\kappa_\lambda$ as a function of $\theta_6$ depends on them. We can see on the left plot, a large $\lambda_5'$ and $\lambda_2'$ lead to a larger variation of $\kappa_\lambda$ as a function of CP-violating phase $\theta_6$.

One interesting aspect to note is that the Yukawa coupling of the 125 GeV Higgs is sensitive to two kinds of phases. On the one hand, they are sensitive to the phase in the Higgs sector, {\it i.e.}\ the phase of $\lambda_6$, on the other hand, they are also sensitive to the phases corresponding to the Yukawa matrices, $\theta_f$'s. On the contrary, the trilinear Higgs coupling is solely sensitive to the phase of $\lambda_6$. Although the sensitivity is quantitatively non-trivial and depends largely on the other parameters of the scalar potential discussed above, it is still possible to make some definitive statements. Depending on whether the effect of CP-violation is seen in the Yukawa sector or in the trilinear self-coupling, one can get an idea of the source of the CP-violation.

\section{Summary and conclusion}
\label{conclusion}

In this work, we explore the CP-violation in a complex-singlet extension of the 2HDM (2HDMS).  We considered the complex singlet extension of a general 2HDM scenario with a specific structure of Yukawa matrices, known as the Yukawa-aligned 2HDM. We have identified the relevant complex phases in the model, which can lead to CP-violation. First, we assume an exact alignment condition, {\it i.e.} the 125 GeV Higgs boson is exactly SM-like. In that case, the CP-violation will be exclusively in the BSM sector. In the exact alignment limit, in a CP-violating (Yukawa-aligned) 2HDMS, 
we found three possible sources of CP-violation: (a) the mixing between the CP-even and CP-odd scalar states, (b) the presence of CP-violating scalar self-couplings, which can appear with or without the aforementioned CP-mixing and (c) the phases of the Yukawa matrices. Notably, the first condition {\it i.e.} the presence of CP-mixing at the tree-level, is absent in the Yukawa-aligned 2HDM, and this is what renders the complex extension its novelty. We showed with our results that, due to the presence of an extra source of CP-violation, parametrized in terms of phase $\theta_{CP}$, the EDM bounds on the model parameter space become significantly relaxed compared to those of 2HDM. We perform a thorough scan and obtain the parameter space allowed by the EDM constraints in our model.

Another novelty of CP-violating 2HDMS is the possibility of incorporating a DM candidate in this model when certain conditions are imposed on the scalar potential. Simultaneous availability of DM and CP-violation gives rise to interesting phenomenological aspects. We investigate the interplay between these two aspects in this model and identify regions of the 
parameter space that are allowed by DM constraints from observed relic density as well as direct and indirect detection experiments. We further impose theoretical as well as experimental constraints on the model parameter space. 

Having identified interesting parameter regions that are allowed by existing data, we investigate the possibility of probing them in future collider experiments. We studied in this context the production of three non-standard scalars at future high-energy $e^+e^-$ colliders, which can give us promising hints regarding CP-violating phases in the BSM sector. 

While it is an interesting scenario that the CP-violation is largely concentrated in the BSM sector and the observed Higgs is mostly CP-even (supported by the data so far), it is imperative that we also look into the CP-violation in the 125 GeV Higgs boson in the context of our model. The major motivation is the ongoing searches and analyses on the experimental front in capturing the CP-nature of the Higgs with increasing precision. In order to study the CP-properties of the 125 GeV Higgs, we had to go beyond the exact alignment condition and investigate the CP-violating Yukawa couplings and the EDM bounds on them. The CP-violation in the Higgs Yukawa coupling has been probed in the $\tau\tau$ interaction at the LHC. We investigated the question of whether the size of the CP-violating phases that can be probed at future colliders, is allowed by the EDM bounds. It turns out that the answer to this question is strongly model-dependent. While the EDM constraint on the CP-phase of 125 GeV Higgs alone is extremely stringent, nevertheless, depending on the parameter space of the model and the presence of additional CP-phases enabling cancellation between possible large contributions, a sizeable CP-phase in the 125 GeV can be allowed by EDM and can be probed at future experiments. In other words, if we observe CP-violation in the Yukawa couplings of the Higgs boson and still do not see any evidence of EDM, it will be most likely that there are further CP-violating phases in nature. Finally, we also study the effect of CP-violation on the trilinear coupling of the Higgs boson. It also turns out that the CP-violating phase of the Higgs has an impact on the trilinear coupling, albeit its nature and extent depend on certain other parameters of the model. Our analysis reveals several notable features of the complex singlet extension of the 2HDM, especially in relation to CP violation and DM. We have further investigated the model-dependent nature of the CP-violating phases and the constraints on them, which directly impact their detectability at future experiments. Therefore, complementary to EFT analyses, we advocate detailed studies of CP-violation and its detection prospects in the context of specific models.

\section*{Acknowledgements}
The authors JL and GMP acknowledge support by the Deutsche Forschungsgemeinschaft (DFG, German Research Foundation) under Germany's Excellence Strategy EXC 2121 "Quantum Universe" - 390833306. The authors acknowledge Shinya Kanemura, Igor P. Ivanov, Georg Weiglein, and Johannes Braathen for their valuable inputs and suggestions.

\newpage
\appendix
\section{Exact formulae of the Barr-Zee type contributions}\label{BarZeeformulae}

Here we list the Barr-Zee contributions to the EDM (CEDM) for a fermion (quark) $d_f^\gamma$, $d_f^Z$ and $d_f^W$ $(d_q^C)$ given in Eqs.~\eqref{edmeq}.
The fermion-loop contributions to the EDM are as follows:
		\small
		\begin{align}
				d_f^V(f')	&=\frac{em_f}{(16\pi^2)^2}8g_{Vff}^vg_{Vf'f'}^v
											Q_{f'}N_C\frac{m_{f'}^2}{v^2}
									\nn\\&\quad\quad
									\times\sum_j^4\int_0^1dz\Bigg\{
												\im[\kappa_f^j]\re[\kappa_{f'}^j]
												\rbra{\frac{1}{z}-2(1-z)}
												+\re[\kappa_f^j]\im[\kappa_{f'}^j]
												\frac{1}{z}
											\Bigg\}
									C^{VH_j}_{f'f'}(z)
				,\\
				d_{f(I_f=-\frac{1}{2})}^W(tb)&=\frac{eg_2^2m_f}{(16\pi^2)^2}N_C
									\int_0^1dz\Bigg\{
											\frac{m_t^2}{v^2}\im[\zeta_f^*\zeta_u]\frac{2-z}{z}
											+\frac{m_b^2}{v^2}\im[\zeta_f^*\zeta_d]
									\Bigg\}[Q_t(1-z)+Q_bz]
									C^{WH^\pm}_{tb}(z)
				,\label{tbup}\\
				d_{f(I_f=+\frac{1}{2})}^W(tb)&=\frac{eg_2^2m_f}{(16\pi^2)^2}N_C
									\int_0^1dz\Bigg\{
											\frac{m_t^2}{v^2}\im[\zeta_f\zeta_u^*]
											+\frac{m_b^2}{v^2}\im[\zeta_f\zeta_d^*]\frac{1+z}{1-z}
									\Bigg\}[Q_t(1-z)+Q_bz]
									C^{WH^\pm}_{tb}(z)
		,\label{tbdown}\end{align}
		\normalsize
\noindent
        where $d_f^V(f')$ denotes the contribution to electric dipole moment (EDM) of fermion $f$, with fermion($f'$) loop and $V=\gamma$, and $Z$ internal leg. $d_{f}^W(tb)$ indicates top-bottom loop contribution to fermion EDM, with $W$ in the internal leg. The internal $W$ leg introduces the weak coupling constant $g_2$ in the expressions~\ref{tbup} and \ref{tbdown}.

The Higgs boson-loop contributions to EDM are
		\begin{align}
				d_f^V(H^\pm)		&=\frac{em_f}{(16\pi^2)^2}4g_{Vff}^v(ig_{H^+H^-V})
									\sum_j^4\im[\kappa_f^j]\frac{g_{H^\pm H^\mp H_j}}{v}
										\int_0^1dz(1-z)
									C^{VH_j}_{H^{^\pm}H^{^\pm}}(z)
				,\\
				d_f^W(H^\pm H)		&=\frac{eg_2^2m_f}{(16\pi^2)^2}\frac{1}{2}
									\sum_j^4\im[\kappa_f^j]\frac{g_{H^{^\pm} H^{^\mp} H_j}}{v}
										\int_0^1dz(1-z)
										C^{WH^\pm}_{H^{^\pm}H_j}(z)
		,\end{align}
\noindent
    where $d_f^V(H^\pm)$ is the contribution to EDM of fermion $f$, via charged Higgs loop, where $V=
    \gamma,Z$ are in the internal leg. $d_f^W(H^\pm H)$ on the other hand implies the charged($H^\pm$) and neutral scalar ($h_i$) loop with $W$ boson in the internal leg.

The gauge-loop contributions to EDM are
		\small
		\begin{align}
				d_f^V(W)		&=\frac{em_f}{(16\pi^2)^2}8g_{Vff}^v(ig_{WWV})\frac{m_W^2}{v^2}
									\sum_j^4\mathcal{R}_{1j}\im[\kappa_f^j]
									\nn\\&\quad\quad
										\times\int_0^1dz\Bigg[
											\cbra{
												\rbra{6-\frac{m_V^2}{m_W^2}}
												+\rbra{1-\frac{m_V^2}{2m_W^2}}\frac{m_{H_j}^2}{m_W^2}
											}\frac{(1-z)}{2}
											-\rbra{4-\frac{m_V^2}{m_W^2}}\frac{1}{z}
										\Bigg]
									C^{VH^0_j}_{WW}(z)
				,\label{gauge1}\\
				d_f^W(WH)	&=\frac{eg_2^2m_f}{(16\pi^2)^2}\frac{1}{2}\frac{m_W^2}{v^2}
									\sum_j^4\mathcal{R}_{1j}\im[\kappa_f^j]
										\int_0^1dz\Bigg\{
										\frac{4-z}{z}-\frac{m_{H^\pm}^2-m_{H_j}^2}{m_W^2}
										\Bigg\}(1-z)
									C^{WH^\pm}_{WH_j}(z)
		,\label{gauge2}\end{align}
		\normalsize

        \noindent
        where $d_f^V(W)$ is the EDM contribution to fermion $f$, through $W$-loop, where the internal leg is $V=\gamma,Z$. $d_f^W(WH)$ is mixed $W$ and neutral scalar loop with $W$ in the internal leg.
        
Finally, the quark-loop contributions to chromo-electric dipole moment (CEDM) are
		\small
		\begin{align}
				d_q^C(q')	&=\frac{m_q}{(16\pi^2)^2}4g_3^3
											\frac{m_{q'}^2}{v^2}
									\sum_j^4\int_0^1dz\Bigg\{
												\im[\kappa_q^j]\re[\kappa_{q'}^j]
												\rbra{\frac{1}{z}-2(1-z)}
												+\re[\kappa_q^j]\im[\kappa_{q'}^j]
												\frac{1}{z}
											\Bigg\}
									C^{gH_j}_{q'q'}(z)
		,\label{cedm}\end{align}
		\normalsize
\noindent
   where $d_q^C(q')$ is the contribution to CEDM of quark (q) with quark ($q'$) loop and gluons in the internal as well as external gauge boson legs. The three interaction vertices of gluon with quarks introduce the third power of the strong coupling constant $g_3^3$ in the expression~\ref{cedm}.

 In all the EDM and CEDM expressions given above, $N_C$ is the color factor.
The coupling constants are given by
\begin{eqnarray}
g_{\gamma ff}^v&=&eQ_f, \nonumber \\
g_{Z ff}^v&=&g_z(I_f/2-Q_fs_W^2), \nonumber \\
g_{H^+ H^- \gamma}&=&-ie, \nonumber \\
g_{H^+ H^- Z}&=&-ig_Z c_{2W}/2, \nonumber \\
g_{H^\pm H^\mp h^0_j}&=&(\lambda_3\mathcal{R}_{1j}+\text{Re}[\lambda_7]\mathcal{R}_{2j}-\text{Im}[\lambda_7]\mathcal{R}_{3j})v + ((\lambda_2'+2\lambda_5')\mathcal{R}_{4j})v_S, \nonumber \\
g_{WWA}&=&-ie, \nonumber \\
g_{WWZ}&=&-ig_Zc_W^2,
\label{definitions}
\end{eqnarray}
where $Q_f$ are the electric charges of the fermion, $g_Z=\sqrt{g_1^2+g_2^2}$, $s_W=\sin\theta_W$, $c_W=\cos\theta_W$ and $c_{2W}=\cos2\theta_W$.
		\begin{align}
				C^{GH}_{XY}(z)	&=C_0\rbra{0,0;m_G^2,m_H^2,\frac{(1-z)m^2_X+zm_Y^2}{z(1-z)}}
		\end{align}
and where $C_0$ is the Passarino-Veltman function\cite{Passarino:1978jh},
		\begin{align}
			C_0(0,0;m_1,m_2,m_3)=\frac{1}{m_1^2-m_2^2}
									\cbra{
										\frac{m_1^2}{m_1^2-m_3^2}\log\rbra{\frac{m_3^2}{m_1^2}}
										-\frac{m_2^2}{m_2^2-m_3^2}\log\rbra{\frac{m_3^2}{m_2^2}}
										}
		.\end{align}
We confirmed the consistency of our results with Refs.~\cite{Leigh:1990kf,Bowser-Chao:1997kjp,Jung:2013hka,Abe:2013qla,Cheung:2014oaa,Cheung:2020ugr}.

\newpage

\section{Interaction basis to Higgs basis transformation}
\label{basis_change}

The basis change equations for the Lagrangian parameters are given as follows.

{\small{\begin{eqnarray*}
 m_{11}^2 &=& {m^I}_{11}^2 c_\beta^2 + {m^I}_{22}^2 s_\beta^2 + \text{Re}[{m^I}_{12}^2] s_{2\beta},\\
 m_{22}^2 &=& {m^I}_{11}^2 s_\beta^2 + {m^I}_{22}^2 c_\beta^2 -\text{Re}[{m^I}_{12}^2] s_{2\beta},\\
 m_{12}^2&=&\mathrm{Re}[{m^I}_{12}^2]c_{2\beta}+({m^I}_{22}^2-{m^I}_{11}^2)s_\beta c_\beta + i \mathrm{Im}[{m^I}_{12}^2],\\
        \lambda_1&=&\lambda_1^I c_\beta^4  +\lambda_2^I s_\beta^4  +2  (\lambda_3^I+\lambda_4^I+\operatorname{Re}[\lambda_5^I])c_\beta^2 s_\beta^2 + 4 (c_\beta^2  \operatorname{Re}[\lambda_6^I] +  s_\beta^2 \operatorname{Re}[\lambda_7^I])s_\beta c_\beta,\\
        \lambda_2&=&\lambda_1^I s_\beta^4  +\lambda_2^I c_\beta^4  +2  (\lambda_3^I+\lambda_4^I+\operatorname{Re}[\lambda_5^I])c_\beta^2 s_\beta^2 - 4 (s_\beta^2  \operatorname{Re}[\lambda_6^I] +  c_\beta^2 \operatorname{Re}[\lambda_7^I] )s_\beta c_\beta,\\
        \lambda_3&=&\lambda_3^I +  \Big({\lambda_1^I + \lambda_2^I} - 2(\lambda_3^I+\lambda_4^I+\operatorname{Re}[\lambda_5^I])\Big)s^2_{ \beta}c^2_\beta-   (\operatorname{Re}[\lambda_6^I]-\operatorname{Re}[\lambda_7^I]  )s_{ 2\beta} c_{2\beta},   \\
        \lambda_4&=&\lambda_4^I +  \Big({\lambda_1^I + \lambda_2^I} - 2(\lambda_3^I+\lambda_4^I+\operatorname{Re}[\lambda_5^I])\Big)s^2_{ \beta}c^2_\beta-   (\operatorname{Re}[\lambda_6^I]-\operatorname{Re}[\lambda_7^I]  )s_{ 2\beta} c_{2\beta},   \\
        \operatorname{Re}[\lambda_5] &=&\operatorname{Re}[\lambda_5^I] +\Big(\lambda_1^I+\lambda_2^I - 2(\lambda_3^I+\lambda_4^I+\operatorname{Re}[\lambda_5^I])\Big)s^2_{\beta}c^2_{\beta}-(\operatorname{Re}[\lambda_6^I]-\operatorname{Re}[\lambda_7^I])s_{ 2\beta} c_{2\beta},\\
        \operatorname{Im}[\lambda_5]&=&\operatorname{Im}[\lambda_5^I] c_{2\beta}-(\operatorname{Im}[\lambda_6^I]-\operatorname{Im}[\lambda_7^I])s_{2\beta},\\
\operatorname{Re}[\lambda_6]&=&\Big(\lambda_2^I s^2_\beta- \lambda_1^I c^2_\beta +( \lambda_3^I + \lambda_4^I + \operatorname{Re}[\lambda_5^I])c_{2\beta} \Big)s_\beta c_\beta +  {(\operatorname{Re}[\lambda_6^I] c^2_\beta+\operatorname{Re}[\lambda_7^I] s^2_\beta)}c_{2\beta} \\
&-& (\operatorname{Re}[\lambda_6^I]-\operatorname{Re}[\lambda_7]^I)2s^2_\beta c^2_\beta,\\
\operatorname{Im}[\lambda_6]&=&  \Big( \operatorname{Im}[\lambda_5^I] s_\beta c_\beta + \operatorname{Im}[\lambda_6^I] c^2_\beta + \operatorname{Im}[\lambda_7^I] s^2_\beta \Big),\\
\operatorname{Re}[\lambda_7]&=&\Big(\lambda_2^I c^2_\beta- \lambda_1^I s^2_\beta -( \lambda_3^I + \lambda_4^I + \operatorname{Re}[\lambda_5^I])c_{2\beta} \Big)s_\beta c_\beta
+ {(\operatorname{Re}[\lambda_6^I] s^2_\beta+\operatorname{Re}[\lambda_7^I] c^2_\beta)}c_{2\beta} \\
&+& (\operatorname{Re}[\lambda_6^I]-\operatorname{Re}[\lambda_7]^I)2s^2_\beta c^2_\beta, \\
\operatorname{Im}[\lambda_7]&=&  -\Big( \operatorname{Im}[\lambda_5^I] s_\beta c_\beta - \operatorname{Im}[\lambda_6^I] s^2_\beta - \operatorname{Im}[\lambda_7^I] c^2_\beta \Big),\\
\lambda_1' &=& \lambda_1'^I c_\beta^2 + \lambda_2'^I s_\beta^2 + \operatorname{Re}[\lambda_6'^I] s_{2\beta},\\
\lambda_2'&=&\lambda_1'^I s_\beta^2 + \lambda_2'^I c_\beta^2 -\operatorname{Re}[\lambda_6'^I] s_{2\beta},\\
\mathrm{Re}[\lambda_4']&=& c_\beta^2 \, \mathrm{Re}[\lambda_4'^I] + s_\beta^2 \, \mathrm{Re}[\lambda_5'^I] + (\mathrm{Re}[\lambda_7'^I] + \mathrm{Re}[\lambda_8'^I]) c_\beta s_\beta,   \\
\mathrm{Im}[\lambda_4']&=& c_\beta^2 \, \mathrm{Im}[\lambda_4'^I] + s_\beta^2 \, \mathrm{Im}[\lambda_5'^I] + (\mathrm{Im}[\lambda_7'^I] + \mathrm{Im}[\lambda_8'^I]) c_\beta s_\beta,\\
\mathrm{Re}[\lambda_5']&=& c_\beta^2 \, \mathrm{Re}[\lambda_5'^I] + s_\beta^2 \, \mathrm{Re}[\lambda_4'^I] - (\mathrm{Re}[\lambda_7'^I] + \mathrm{Re}[\lambda_8'^I]) c_\beta s_\beta,  \\
\mathrm{Im}[\lambda_5']&=& c_\beta^2 \, \mathrm{Im}[\lambda_5'^I] + s_\beta^2 \, \mathrm{Im}[\lambda_4'^I] - (\mathrm{Im}[\lambda_7'^I] + \mathrm{Im}[\lambda_8'^I]) c_\beta s_\beta,   \\
\mathrm{Re}[\lambda_6']&=&(\lambda_2'^I-\lambda_1'^I)s_\beta c_\beta +\operatorname{Re}[\lambda_6'^I] c_{2\beta}, \\
\mathrm{Im}[\lambda_6']&=&\operatorname{Im}[\lambda_6'^I], \\
\mathrm{Re}[\lambda_7']&=& c_\beta^2 \mathrm{Re}[{\lambda_7'}^I] + c_\beta s_\beta (\mathrm{Re}[{\lambda_5'}^I] - \mathrm{Re}[\lambda_4'^I]) - s_\beta^2 \mathrm{Re}[{\lambda_8'}^I],  \\
\mathrm{Im}[\lambda_7']&=& c_\beta^2 \mathrm{Im}[{\lambda_7'}^I] + c_\beta s_\beta (\mathrm{Im}[{\lambda_5'}^I] - \mathrm{Re}[\lambda_4'^I]) - s_\beta^2 \mathrm{Im}[{\lambda_8'}^I],\\
\mathrm{Re}[\lambda_8']&=& c_\beta^2 \mathrm{Re}[{\lambda_8'}^I] + c_\beta s_\beta (\mathrm{Re}[{\lambda_5'}^I] - \mathrm{Re}[\lambda_4'^I]) - s_\beta^2 \mathrm{Re}[{\lambda_7'}^I], \\
\mathrm{Im}[\lambda_8']&=& c_\beta^2 \mathrm{Im}[{\lambda_8'}^I] + c_\beta s_\beta (\mathrm{Im}[{\lambda_5'}^I] - \mathrm{Re}[\lambda_4'^I]) - s_\beta^2 \mathrm{Im}[{\lambda_7'}^I]. 
\end{eqnarray*}}}

\noindent
The parameters on the left-hand side of the equations are given in the Higgs basis, and the parameters with a superscript $I$ on the right-hand side of the equations are in the interaction basis. 

\section{Interaction terms between DM $a_S$ and the visible scalars states in the Higgs basis}
\label{dm_interactions}

\noindent
Coefficients of the bilinear interaction terms involving DM candidate $a_S$ and other scalar degrees of freedom are as follows.

\begin{eqnarray*}
   h_1 a_S &=& 2 \text{Im}[\lambda_4'] v v_S, \\
   h_2 a_S &=&  \text{Im} [\lambda_7'+ \lambda_8'] v v_S, \\
   a_2 a_S &=&  \text{Re} [\lambda_7'- \lambda_8'] v v_S, \\
   h_S a_S &=& (\text{Im} [m_S'^2] + \text{Im}[\lambda_4'] v^2 + \frac{1}{2} \text{Im} [\lambda_1''+ 2 \lambda_2''] v_S^2).\\
   \end{eqnarray*}

\noindent
Coefficients of the trilinear interaction terms involving $a_S$ and other scalars are as follows.

\begin{eqnarray*}
   h_1 h_1 a_S &=& 2 \text{Im}[\lambda_4'] v_S, \\
   h_1 h_2 a_S &=&  \text{Im} [\lambda_7'+ \lambda_8'] v_S, \\
   h_1 a_2 a_S &=&  \text{Re} [\lambda_7'- \lambda_8'] v_S, \\
   h_1 h_S a_S &=& 2 \text{Im} [\lambda_4'] v_S, \\
   h_2 h_2 a_S &=&  \text{Im} [\lambda_5'] v_S, \\
   h_2 a_2 a_S &=& 0, \\
   h_2 h_S a_S &=&  \text{Im} [\lambda_7'+ \lambda_8'] v, \\
   a_2 a_2 a_S &=&  \text{Im} [\lambda_5'] v_S, \\
   a_2 h_S a_S &=&   \text{Re} [\lambda_7'- \lambda_8'] v, \\
   h_S h_S a_S  &=& \frac{1}{4} \text{Im} [\lambda_1''+ 2 \lambda_2''] v_S. 
   \end{eqnarray*}

\noindent
Coefficients of the quartic interaction terms involving $a_S$ and other scalars are given below.

\begin{eqnarray*}
h_1 h_1 h_1 a_S &=& 0, \\
h_1 h_1 h_2 a_S &=& 0, \\
h_1 h_1 a_2 a_S &=& 0, \\
h_1 h_1 h_S a_S &=& \text{Im}[\lambda_4'], \\
h_1 h_2 h_2 a_S &=& 0, \\
h_1 h_2 a_2 a_S &=& 0, \\
h_1 h_2 h_S a_S &=&  \text{Im} [\lambda_7'+ \lambda_8'], \\
h_1 a_2 a_2 a_S &=& 0, \\
h_1 a_2 h_S a_S &=&  \text{Re} [\lambda_7'- \lambda_8'], \\
h_1 h_S h_S a_S &=& 0, \\
h_2 h_2 h_2 a_S &=& 0, \\
h_2 h_2 a_2 a_S &=& 0, \\
h_2 h_2 h_S a_S &=& \text{Im} [\lambda_5'], \\
h_2 a_2 a_2 a_S &=& 0, \\
h_2 a_2 h_S a_S &=& 0, \\
h_2 h_S h_S a_S &=& 0, \\
a_2 a_2 a_2 a_S &=& 0, \\
a_2 a_2 h_S a_S &=&  \text{Im} [\lambda_5'], \\
a_2 h_S h_S a_S &=& 0, \\
h_S h_S h_S a_S &=& \frac{1}{12} \text{Im} [\lambda_1''+ 2 \lambda_2''].
\end{eqnarray*}

\bibliographystyle{JHEP}
\bibliography{ref1}

\end{document}